\documentclass[twocolumn,showpacs,preprintnumbers,floatfix,amsmath]{revtex4}
\usepackage[dvips]{graphicx}
\usepackage{dcolumn}
\usepackage{bm}
\usepackage{epsf}

\begin{document}

\title{Bump formations in binary attractor neural network}

\author{
Kostadin Koroutchev
}
\email{k.koroutchev@uam.es}
\affiliation{
Depto. de Ingenier\'{\i}a Inform\'{a}tica\\
Universidad Aut\'{o}noma de Madrid, 28049 Madrid, Spain
}
\altaffiliation[Also at]{
Inst. for Communications and Computer Systems, Bulgarian Academy of Sciences, 
1113, 
Sofia, Bulgaria}

\author{Elka Korutcheva }
\affiliation{
Depto. de F\'{\i}sica Fundamental, Universidad Nacional de
Educaci\'on a Distancia,\\
c/ Senda del Rey 9,28080 Madrid, Spain
}
\altaffiliation[Also at]{
G.Nadjakov Inst.Solid State
  Physics, Bulgarian Academy of Sciences, 1784 Sofia, Bulgaria
}


\begin{abstract}
This paper investigates the conditions for the formation of local bumps
in the activity of binary attractor neural networks with 
spatially dependent connectivity.
We show that
these formations are observed when asymmetry between the activity during the
retrieval and learning is imposed.
Analytical approximation for the order parameters is derived.
The corresponding phase diagram
shows a relatively large and stable region,
where this effect is observed, although the critical storage
and the information capacities
drastically decrease inside that region.
We demonstrate that 
the stability of the network, when starting from the bump formation, is larger 
than the stability when starting even from the whole pattern.
Finally, we show a very good agreement between the analytical results and
the simulations performed for different topologies of the network.
\end{abstract}


\maketitle

\section{Introduction}
The bump formations in recurrent neural networks have been recently 
reported
in several investigations concerning linear-threshold units \cite{AleYasser1,AleYasser2}, binary units \cite{EK1,EK2}, Small-World networks of Integrate
and Fire neurons \cite{qbio} and in a variety of spatially distributed neural
networks models with excitatory and inhibitory couplings between the cells
\cite{Rubin,Brunel}.

As has been shown, the localized retrieval is due to the
short-range connectivity of the networks and could explain the behavior in 
structures of biological relevance as the neocortex,  where the
probability of connections decreases with the distance \cite{Braitenberg}.

In the case of linear-threshold neural network model, the signal-to-noise
analysis has been recently adapted \cite{AleYasser1,AleYasser2} to spatially
organized networks and has shown that the retrieval states of the connected
network have non-uniform activity profiles when the connections are
short-range enough, even without any spatial structure of the stored
patterns. The increase of the gain or the saturation level of neurons enhance
the level of localization of the retrieval states and do not lead to a drastic
decrease of the storage capacity in these networks even for very localized
solutions \cite{AleYasser2}.

An interesting investigation of the spontaneous activity bumps in Small-World
networks (SW) \cite{SW1,SW2} of Integrate-and Fire neurons \cite{qbio}, has
recently shown that the network retrieves when its connectivity is close to 
the random and displays localized bumps
of activity, when its connectivity is close to the ordered. The two regimes
are mutually exclusive in the range of the parameter governing the proportion
of the long-range connections on the SW topology of Integrate-and-Fire
network, while the two behaviors coexist in the case of linear-threshold and
smoothly saturated units. Moreover, it has been stated that the transition
between localization and retrieval regimes, in the case of SW networks of
Integrate-and-Fire neurons, can occur at a degree of randomness beyond the SW
regime and it is not related, in general, to the SW properties of the network.

The bump formations have been also investigated in spatially distributed
neural network models with excitatory and inhibitory synaptic couplings.
Recently it has been shown \cite{Rubin} that a neural network only with
excitatory couplings can exhibit localized activity from an initial transient
synchrony of a localized group of cells, followed by desynchronized activity
within the group. This activity may grow or may be reduced when depression
with a given size of the frequency  of the inputs is introduced. It is also
very sensitive to the initial conditions and the range of the parameters of
the network.

The result of bump formations have been recently reported by us, \cite{EK1,EK2}
in the case of binary Hebb model for associative network. We stated out that
these spatially asymmetric retrieval states (SAS) can be observed if an
asymmetry between  the learning and the retrieval states is imposed. This
means that the network is constrained to have a different activity compared to
that induced by the patterns.

In the present investigation we 
regard a symmetric and 
distance dependent connectivity
for all neurons within an attractor neural network (NN) of
Hebbian type formed  by $N$ binary
neurons $\{S_i\}, S_i\in \{-1,1\}, i=1,...,N$,
storing $p$ binary patterns $\eta_i^{\mu}, \mu\in \{ 1...p\}$.
The connectivity between the neurons is symmetric 
$c_{ij}=c_{ji}\in\{0,1\}, c_{ii}=0$, where $c_{ij}=1$
means that neurons $i$ and $j$ are connected.

We are interested only on connectivities in which the fluctuations between
the individual connectivity are small, e.g.
$\forall_{i} \sum_j c_{ij}\approx c N$,
where $c$ is the mean connectivity.

The learned patterns are drawn from the following distribution:
\[
P(\eta_i^\mu)=\frac{1+a}{2}\delta(\eta_i^\mu-1)+\frac{1-a}{2}
\delta(\eta_i^\mu+1),
\]
where the parameter $a$ is the sparsity of the code \cite{feigelman,Bolle}.

We study the Hopfield model \cite{Hopfield}:
\begin{equation}
H = \frac{1}{N}\sum_{ij} J_{ij} S_i S_j,
\end{equation}
with Hebbian learning rule \cite{Hebb}
\begin{equation}
J_{ij} = \frac{1}{c}\sum_{\mu=1}^p c_{ij}
(\eta_{i}^{\mu} - a) (\eta_{j}^{\mu} - a) .
\end {equation}
Further in this article we will work in terms of variables
$\xi_{i}^{\mu} \equiv \eta_{i}^{\mu} - a$.

In the case of binary network and symmetrically distributed patterns,
the only asymmetry between the retrieval and the learning states
that can be imposed, independent on the position of the
neurons, is the total number of the neurons in a given state. Having in
mind that there are only two possible states, this condition leads to a
condition on the mean activity of the network, which is introduced
by adding
an extra term $H_a$ to the Hamiltonian
\[
H_a = N R (\sum_i S_i/N - a).
\]
This term favors states with
lower total activity $\sum_i S_i$ that is equivalent to
decrease the number of active neurons, creating
asymmetry between the learning and the retrieval states.

For $R=0$, this model has been intensively studied 
\cite{feigelman,Bolle,Amit1} and shows no evidence of spatial
asymmetric states.

For $R>0$, the energy of the system increases with the number of
active neurons ($S_i=1$) and the term $H_a$ tends to limit the number of
active neurons below $aN$.

In the present work we show explicitly that the
constraint on the network level of activity
is a sufficient condition for observation of spatially
asymmetric states in binary Hebb neural network.
Similar observations have been reported in the case of linear-threshold
  network \cite{AleYasser1}, where in order to observe bump formations, one
  has to constrain the activity of the network. The same is true in the case
  of smoothly saturating and binary networks \cite{AleYasser2}, when the
  highest activity level, they can achieve, is above the maximum activity of 
the units in the stored pattern.

In the present paper we discuss in details the binary neural network 
model,
giving  a complete analytical derivation of its properties. 
We explain the minimal conditions for the formations of spatially asymmetric 
states in a recurrent network with metrically organized connectivity. 

We compare the
results for three different topologies of the network and show interesting
multi-bump formations in the case of the SW-like topology. 
Further extension of the
analysis to the information, carried by the network, reveals that the 
retrieved states can be well 
recovered by a limited amount of information. 
We show that the stability of the network's retrieval, when starting from the 
bump, is larger than the stability, when starting even from the whole pattern.
This could be useful in
practical retrieval tasks with minimal information transmitted.

The paper is organized as follows:
In section {\em Analytical analysis} and using replica theory, 
we present our detailed mean-field analysis in the case 
of distance-dependent connectivities.
We derive the
equations for the corresponding order parameters (OP) for finite and
zero temperatures. In section {\em Simulations} we present the results of the
simulations,
done for the case of different topologies of the network, and we compare them
with the analytical results. In section {\em Discussion} we focus on
the stability of the phase diagram of the network and especially on the
dependence of the SAS region on the parameters of the model. We discuss
the behavior of the critical storage capacity when bump formations are present,
as well as their effect on the information capacity of the network.
The conclusion, drawn in the last part, shows that only the
asymmetry between the learning and the retrieval states is
sufficient to observe spatially asymmetric states.

\section{Analytical analysis}

For the analytical analysis of the SAS states, we
consider the decomposition of the connectivity matrix
$c_{ij}$ by its eigenvectors $a_i^{(k)}$:
\[
\label{connectiv}
c_{ij}=\sum_k \lambda_k a_i^{(k)} a_j^{(k)}, \,\,\, \sum_i 
a_i^{(k)} a_i^{(l)} =\delta_{kl},
\]
where $\lambda_k$ are the corresponding (positive) eigenvalues.
For convenience
we denote $b_i^k\equiv a_i^{(k)}\sqrt{\lambda_k}$, having
\[
c_{ij}=\sum_k b_i^{k} b_j^{k}.
\]

\begin{figure}[ht!]
\begin{center}
\begin{minipage}{5.70cm}
\epsfxsize 5.7cm 
\epsfbox{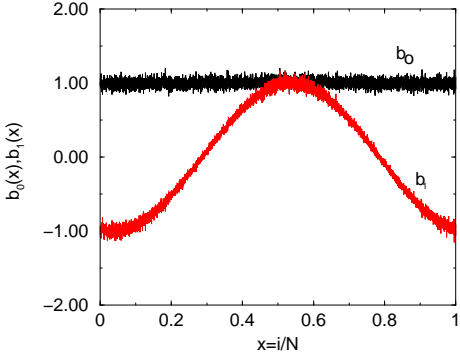}
\end{minipage}
\caption{\small
(Color online) 
The components of the first and the second eigenvectors of the 
connectivity
matrix for $N=6400$, $c=320/N$, $\lambda_1=319.8$, $\lambda_2=285.4$.
The first eigenvector has
constant components, the second one -
sine-like components.
}
\label{fig4}
\end{center}
\end{figure}

We will assume that $a_i^{(k)}$ are ordered by their
eigenvalues in decreasing order, e.g. for
$k>l \Rightarrow \lambda_k\leq \lambda_l$.
To get some intuition of what $a_j^{(k)}$
 look like, we plot in Fig. \ref{fig4}
the first two eigenvectors.

For a wide variety of connectivities, the first
three eigenvectors approximate:
\begin{equation}
\label{a0}
a_k^{(0)}=\sqrt{{1}/{N}},
\end{equation}
\begin{equation}
\label{a1}
a_k^{(1)}=\sqrt{{2}/{N}}\cos(2\pi |k-k_0|/N)
\end{equation}
and
\begin{equation}
\label{a2}
a_k^{(2)}=\sqrt{{2}/{N}}\sin(2\pi |k-k_0|/N).
\end{equation}

Following the classical analysis of Amit et al. \cite{Amit1}, 
we use Bogolyubov's method of quasi averages \cite{Bogol} 
to have into account a finite numbers of overlaps that condense 
macroscopically. To this aim we introduce an external field, 
conjugate to a finite number of patterns $\{\xi_{i}^{\nu}\}, \nu=1,2,...,s$, 
adding to the Hamiltonian the term
$H_h = \sum_{\nu=1}^s h^{\nu} \sum_i \xi_{i}^{\nu}S_i$.

In order to impose some asymmetry in the neural network's states, 
we also add the term $H_a=N R \left(\sum_i S_i/N -a \right)$.

The whole Hamiltonian, we are studying, is:
\begin{eqnarray}
\label{Hamiltonian}
H = \frac{1}{cN}\sum_{ij\mu} S_i \xi_i^\mu c_{ij} \xi_j^\mu S_j + 
\sum_{\nu=1}^s h^{\nu} \sum_i \xi_{i}^{\nu}S_i \nonumber\\
+ N R (\sum_i S_i/N-a).
\end{eqnarray}

By using the ``replica method'' \cite{mezard}, 
for the averaged free energy per spin we get:
\begin{equation}
\label{fenergy}
f= - \lim_{n\rightarrow 0} \lim_{N\rightarrow \infty} \frac{1}{\beta n N} 
({\langle\langle}Z^{n}{\rangle\rangle} -1),
\end{equation}
where ${\langle\langle}...{\rangle\rangle}$ stands for the average over 
the pattern distribution $P(\xi_{i}^{\mu})$, $n$ is the number of the 
replicas, which are later taken to zero and $\beta$ is the inverse temperature.
\begin{widetext}
The replicated partition function is
\begin{eqnarray}
{\langle\langle}Z^n{\rangle\rangle} =
{\biggl\langle\biggl\langle}
  Tr_{S^{\rho}}
    \exp\biggl[
      \frac{\beta}{2cN}
        \sum_{ij\mu\rho}
             (\xi_{i}^{\mu}S_{i}^{\rho})c_{ij}
             (\xi_{j}^{\mu}S_{j}^{\rho})
         -\frac{1}{2c}\beta pn + \nonumber\\
      \beta \sum_{\nu} h^{\nu} \sum_{i,\rho} \xi_{i}^{\nu} S_{i}^{\rho}
      - \sum_{\rho}
          \beta N R
          (\sum_{i} S_{i}^{\rho}/N -a)
    \biggl]
{\biggl\rangle\biggl\rangle} .
\end{eqnarray}
The decomposition of the sites, by using an expansion of the 
connectivity matrix $c_{ij}$ over its eigenvalues
$\lambda^{l}, l=1,...,M$ and its eigenvectors $a_{i}^{l}$,  gives:
\begin{eqnarray}
{\langle\langle}Z^n{\rangle\rangle}=
e^{-\beta p n/ 2 c}
{\biggl\langle\biggl\langle}
 Tr_{S^{\rho}}
  \exp\biggl[
    \frac{\beta}{2cN}
      \sum_{\mu\rho l}\sum_{ij}
            (\xi_{i}^{\mu} S_{i}^{\rho} b_{i}^{l})
            (\xi_{j}^{\mu} S_{j}^{\rho} b_{j}^{l})+ \nonumber\\
      \beta \sum_{\nu} h^{\nu} \sum_{i \rho} \xi_{i}^{\nu} S_{i}^{\rho} -
      \sum_{\rho}\beta R N (\sum_{i} S_{i}^{\rho}/N -a)
  \biggl]
{\biggl\rangle\biggl\rangle} .
\end{eqnarray}
Introducing variables $m_{\rho l}^{\mu}$ at each replica $\rho$, 
each configuration and each eigenvalue, we get:
\begin{eqnarray}
{\langle\langle}Z^n{\rangle\rangle}=
e^{-\beta p n/ 2c + \beta Ra N} \nonumber\\
{\biggl\langle\biggl\langle}
 Tr_{S^{\rho}}
  \int \prod_{\mu l \rho}
    \frac{dm_{l}^{\mu}}{\sqrt{2 \pi}}
    \exp \beta cN
      \left(
       -\frac{1}{2} \sum_{\mu \rho l}(m_{\rho l}^{\mu})^2 +
       \sum_{\mu \rho l}m_{\rho l}^{\mu}\frac{1}{cN}
          \sum_{i}\xi_{i}^{\mu} S_{i}^{\rho} b_{i}^{l}
      \right) \nonumber\\
    \exp{\beta cN
      \left(
       -\frac{1}{2} \sum_{\nu \rho l}(m_{\rho l}^{\nu})^2 +
       \sum_{\nu \rho l}m_{\rho l}^{\nu}\frac{1}{cN}
         \sum_{i}\xi_{i}^{\nu} S_{i}^{\rho} b_{i}^{l}
         + h^{\nu} \frac{1}{N}
           \sum_{i} (\xi_{i}^{\nu} S_{i}^{\rho} + R S_{i}^{\rho})
      \right)}
{\biggl\rangle\biggl\rangle},
\end{eqnarray}
\end{widetext}
where we have split the sums over the first $s$-patterns and 
the remaining (infinite) $p-s$ ones.

The ``condensed'' order parameters $m_{\rho l}^{\nu}$ 
have a clear physical meaning as an overlap
between the pattern $\nu$ and the state of the neuron in the replica $\rho$
of the system, modulated by the $l$-th eigenvalue of the connectivity matrix:
\begin{equation}
\label{m0}
m_{\rho l}^{\nu} = \frac{1}{cN} \sum_{i}\xi_{i}^{\nu} S_{i}^{\rho} b_{i}^{l} .
\end{equation}
In the present investigation, the main physical hypothesis is
that only a finite number, $k$, of the
order parameters $m_{\rho,l}^\nu,\ l=1,...,k$
are macroscopically different from zero.
As we can see later in this paper the simulations
also confirm this hypothesis.
This avoids the introduction of infinite number of parameters,
dependent on the state of the system, that would make impossible
the analytical solution of the problem.

After taking the averages over the pattern, for the first term we get:
\begin{widetext}
\begin{equation}
\label{expan}
I = \exp \beta \left(-\frac{1}{2} \sum_{\mu \rho l} (m_{\rho l}^{\mu})^2 
+ \frac{\beta(1-a^2)}{2cN} \sum_{\rho \sigma l k i \mu} m_{\rho l}^{\mu}
 m_{\sigma k}^{\mu} S_{i}^{\rho} S_{i}^{\sigma} b_{i}^{l} b_{i}^{k}\right) .
\end{equation}
The integration of the last expression over the OP $m_{\rho l}^{\mu}$ gives:
\begin{eqnarray}
\label{first_part}
\int \prod_{\mu \rho l}\frac{dm_{\rho l}^{\mu}}{\sqrt{2 \pi}} I = \nonumber\\
\int \prod_{\rho\sigma l k} dq_{\rho \sigma}^{lk}
  \exp\left(-\frac{p}{2} Tr \ln [A_{\rho\sigma}^{lk}]\right)
  \prod_{\rho\sigma l k}
    \delta(q_{\rho \sigma}^{l k} -
      \frac{1}{cN}
      \sum_{i} S_{i}^{\rho} S_{i}^{\sigma} b_{i}^{l} b_{i}^{k}) .
\end{eqnarray}
\end{widetext}
The expression within the $\delta$-function can be regarded as a definition
of the OP $q_{\rho \sigma}^{lk}$.

Now let us suppose that the order parameter
$q_{\rho \sigma}^{l k}$ can split as a product of two terms: 
one, which only depends on the replica's and the 
eigenvalue's indexes and another one
that introduces the spatial dependence of the distribution of the eigenvalues
and eigenvectors:
$q_{\rho \sigma}^{l k} \equiv
q_{\rho \sigma}^{l} \sum_{i}b_{i}^{l} b_{i}^{k}/cN =  
q_{\rho \sigma}^{l} \delta_{lk}(\lambda_l/cN)$.
In this way
we can keep only one index for  the OP $q_{\rho \sigma}^l$ when 
taking into account the
spatial distribution of the matrix of interactions.

Introducing the parameter $r_{\rho \sigma}^{l}$, conjugate to
$q_{\rho \sigma}^{l}$, for the last expression we obtain:
\begin{widetext}

\begin{eqnarray}
\int \prod_{\mu \rho l}\frac{dm_{\rho l}^{\mu}}{\sqrt{2 \pi}} I =
\int
  \prod_{\rho\sigma l}
    dq_{\rho \sigma}^{l}
  \prod_{\rho\sigma l}
    dr_{\rho \sigma}^{l}
  \exp
   \left(
    -\frac{p}{2} Tr \ln [A_{\rho\sigma}^{l}]
   \right)\nonumber\\
  \exp cN
    \left(
    -\frac{1}{2} \alpha \beta^2(1-a^2)
       \sum_{\rho \sigma l} r_{\rho \sigma}^{l} q_{\rho \sigma}^{l}
    +\frac{1}{2cN} \alpha \beta^2 (1-a^2)
       \sum_{i \rho \sigma l}
         r_{\rho \sigma}^{l}
         S_{i}^{\rho} S_{i}^{\sigma}
         b_{i}^{l} b_{i}^{l}
    \right),
\end{eqnarray}
where the parameter $\alpha \equiv {p}/{N}$ is the storage
capacity of the network.

The matrix $A_{\rho\sigma}^{l}$ is
\begin{equation}
\label{thematrix}
A_{\rho\sigma}^{l}=\delta_{\rho \sigma}\left(1 -
\beta (1-a^2)\mu_l(1-q_l)\right) -
\beta (1-a^2) q_l \mu_l
\end{equation}
and $\mu_l = \lambda_l /cN$.

For the replicated partition function
${\langle\langle}Z^{n}{\rangle\rangle}$,
after taking the limit
  $h^{\nu} \rightarrow 0$, we obtain:
\begin{eqnarray}
{\langle\langle}Z^n{\rangle\rangle}
  & =
   & e^{-\beta p n(1-a^2)/2c + \beta R a n N}
     \int \prod_{\nu}
       dm_{\rho l}^{\nu}
     \int \prod_{\rho \sigma l}
       dq_{\rho \sigma}^{lk} dr_{\rho \sigma}^{l} \nonumber\\
  && \exp cN
      \left(
        -\frac{\beta}{2} \sum_{\nu \rho l} (m_{\rho l}^{\nu})^2
        - \frac{1}{2} Tr \ln [A_{\rho\sigma}^{l}]
        - \frac{1}{2} \alpha \beta^2(1-a^2)
            \sum _{\rho\neq\sigma,l}r_{\rho \sigma}^{l}q_{\rho \sigma}^{l}
      \right) \nonumber\\
  && {\biggl\langle\biggl\langle}
      Tr_{S^{\rho}}
       \exp cN
        \biggl[
          \frac{1}{2cN} \alpha \beta^2
           \sum_{i\rho\sigma l k}
             r_{\rho \sigma}^{l}
             S_{i}^{\rho}S_{i}^{\sigma} b_{i}^{l} b_{i}^{l} + \nonumber\\
  && \beta
     \sum_{\nu \rho l}
       m_{\rho l}^{\nu}\frac{1}{cN}
        \sum_{i}\xi_{i}^{\nu} S_{i}^{\rho} b_{i}^{l} +
     \beta R
     \sum_{i}S_{i}^{\rho}
     \biggl]
{\biggl\rangle\biggl\rangle} .
\end{eqnarray}

Supposing Replica Symmetry (RS) ansatz, i.e., $m_{\rho}^{\nu} = m^{\nu}$ for
any replica index $\rho$ and $q_{\rho\sigma}^{l} = q^{l}, r_{\rho\sigma}^{l}
= r_{l}$ for $\rho \neq \sigma$, for the free energy, Eq.(\ref{fenergy})
we obtain:
\begin{eqnarray}\label{ref17}
f= \frac{\alpha(1-a^2)}{2c}
   + Ra
   + \frac{\alpha}{2\beta n} Tr \ln[A^{lk}]
   + \frac{1}{2}
      \sum_{\nu l} (m_{l}^{\nu})^2
   - \frac{\alpha \beta(1-a^2)}{2} \sum_{lk} r_{l} q^{l} - \nonumber\\
     \frac{1}{n\beta}
     {\biggl\langle\biggl\langle}
      \ln Tr_{S^{\rho}}
       \exp \biggl[
         \frac{1}{2cN}\alpha\beta^2 (1-a^2)
         \sum_{i \rho \sigma l }
  r_{\rho \sigma}^{l}\overline{S_{i}^{\rho}S_{i}^{\sigma}b_{i}^{l}b_{i}^{l}}
           + \nonumber\\
         \beta
         \sum_{\nu \rho l}
           m_{l}^{\nu}
           \frac{1}{cN} \sum_{i}\xi_{i}^{\nu} S_{i}^{\rho} b_{i}^{l}
       + \beta R \sum_{i} S_{i}^{\rho}
       \biggl]
     {\biggl\rangle\biggl\rangle} .
\end{eqnarray}
\end{widetext}
The over-line in the last expression has its clear physical meaning as an
average over the spatial distribution
$\overline{(.)_i} = \frac{1}{cN}\sum_{i} (.)$.

As a next step, let us suppose that the average over a finite number of
patters $\xi^\nu$ can be self-averaged \cite{Amit1}. In our case this is
expressed by the following decomposition
$\overline{S_{i}^{\rho}S_{i}^{\sigma}b_{i}^{l}b_{i}^{l}} \approx
S^{\rho} S^{\sigma} \overline{(b_{i}^{l})^2}$,
which is reasonable to assume as a first approximation by taking into
account the effect of the spatial distribution.
This approximation permits to
complete the analytical analysis.

After taking the trace $Tr_{S^\rho}$ and
the limit $n \rightarrow 0$, in Eq. (\ref{ref17})
and using
the saddle point method,
we end up with
the following expression for the free energy: 
\begin{widetext}
\begin{eqnarray}
f& = &\frac{1}{2c}\alpha(1-a^2) + \frac{1}{2} \sum_k (m_{k})^2
    - \frac{\alpha\beta(1-a^2)}{2} \sum_{k} r_k q_{k}
    + \frac{\alpha\beta(1-a^2)}{2} \sum_k \mu_k r_k +\nonumber\\
&+&
\frac{\alpha}{2\beta}\sum_k [\ln(1-\beta(1-a^2) \mu_k + \beta(1-a^2)q_k) -\\
&-&\beta(1-a^2) q_k(1-\beta(1-a^2)\mu_k + \beta(1-a^2) q_k)^{-1}]-\nonumber\\
\nonumber\\
&-&
    \frac{1}{\beta}
     \int{\frac{dz e^{-z^2/2}}{\sqrt{2\pi}}}
       \overline{
         \ln 2 \cosh \beta \left(
            z \sqrt{ \alpha (1-a^2) \sum_l r_l b_{i}^{l} b_{i}^{l}}
       + \sum_l m_{l} \xi_i b_{i}^{l}
       + R b_i^0\right)} .\nonumber
\end{eqnarray}

The equations for the OP $r_k$, $m_k$ and $q_k$  are respectively:

\begin{eqnarray}
\label{r}
  r_k=\frac{q_k (1-a^2)}{\left(1-\beta (1-a^2)(\mu_k-q_k)\right)^2} ,
\end{eqnarray}

\begin{eqnarray}
\label{m}
m_k = \int \frac{d z e^{-z^2/2}}{\sqrt{2\pi}}
       \overline{
          \xi_i b_i^k\tanh \beta \left(
            z \sqrt{ \alpha (1-a^2) \sum_l r_l b_{i}^{l}b_{i}^{l} }
       + \sum_{l} m_{l} \xi_i b^l_i
       + R b_i^0\right)}
\end{eqnarray}
and
\begin{eqnarray}
\label{q}
q_k =  \int\frac{dz e^{-z^2/2}}{\sqrt{2\pi}}
       \overline{
          (b_i^k)^2 \tanh^2 \beta \left(
            z \sqrt{\alpha (1-a^2)\sum_{l} r_l b_{i}^{l} b_{i}^{l}}
       + \sum_{l}m_{l} \xi_i b^l_i
       + R b_i^0\right)} .
\end{eqnarray}

The following analysis refers to the case $T=0$.
Keeping $C_k\equiv\beta (\mu_k-q_k)$ finite and limiting the above
system
only to the first two $m_k$-s,
the Eqs. (\ref{r}-\ref{q}) read:
\begin{eqnarray}
\label{T=0}
  m_0&=& \frac{1-a^2}{4\pi}\int_{-\pi}^{\pi} g(\phi) d\phi\\
  m_1&=& \sqrt{2\mu_1}\frac{1-a^2}{4\pi}\int_{-\pi}^{\pi}g(\phi)
\sin\phi\ d\phi\\
  C_0&=& \frac{1}{2\pi}\int_{-\pi}^{\pi} g_c(\phi) d\phi\\
  C_1&=& \frac{\mu_1}{\pi} \int_{-\pi}^{\pi} g_c(\phi) \sin^2\phi\ d\phi\\
  r_k&=& \frac{\mu_k (1-a^2)}{[1-(1-a^2)C_k]^2}\label{tzeron} .
\end{eqnarray}

Here
\begin{eqnarray}
g(\phi)&=&{\rm erf}(x_1)+{\rm erf}(x_2)\\
g_c(\phi)&=&[(1+a)e^{-(x_1)^2}+(1-a)e^{-(x_2)^2}]/[{\sqrt{\pi}y}]\\
x_1&=&{[(1-a)(m_0+m_1\sqrt{2 \mu_1}\sin\phi)+R]}/{y}\\
x_2&=&{[(1+a)(m_0+m_1\sqrt{2 \mu_1}\sin\phi)-R]}/{y}\\
y&=&\sqrt{2\alpha (1-a^2) (r_0+2\mu_1 (r_1-1+a^2)\sin^2\phi)}\label{tzero1}.
\end{eqnarray}
\begin{figure}[t]
\begin{center}
$\begin{array}{cc}
\begin{minipage}{5cm}
\epsfxsize 5cm 
\epsfbox{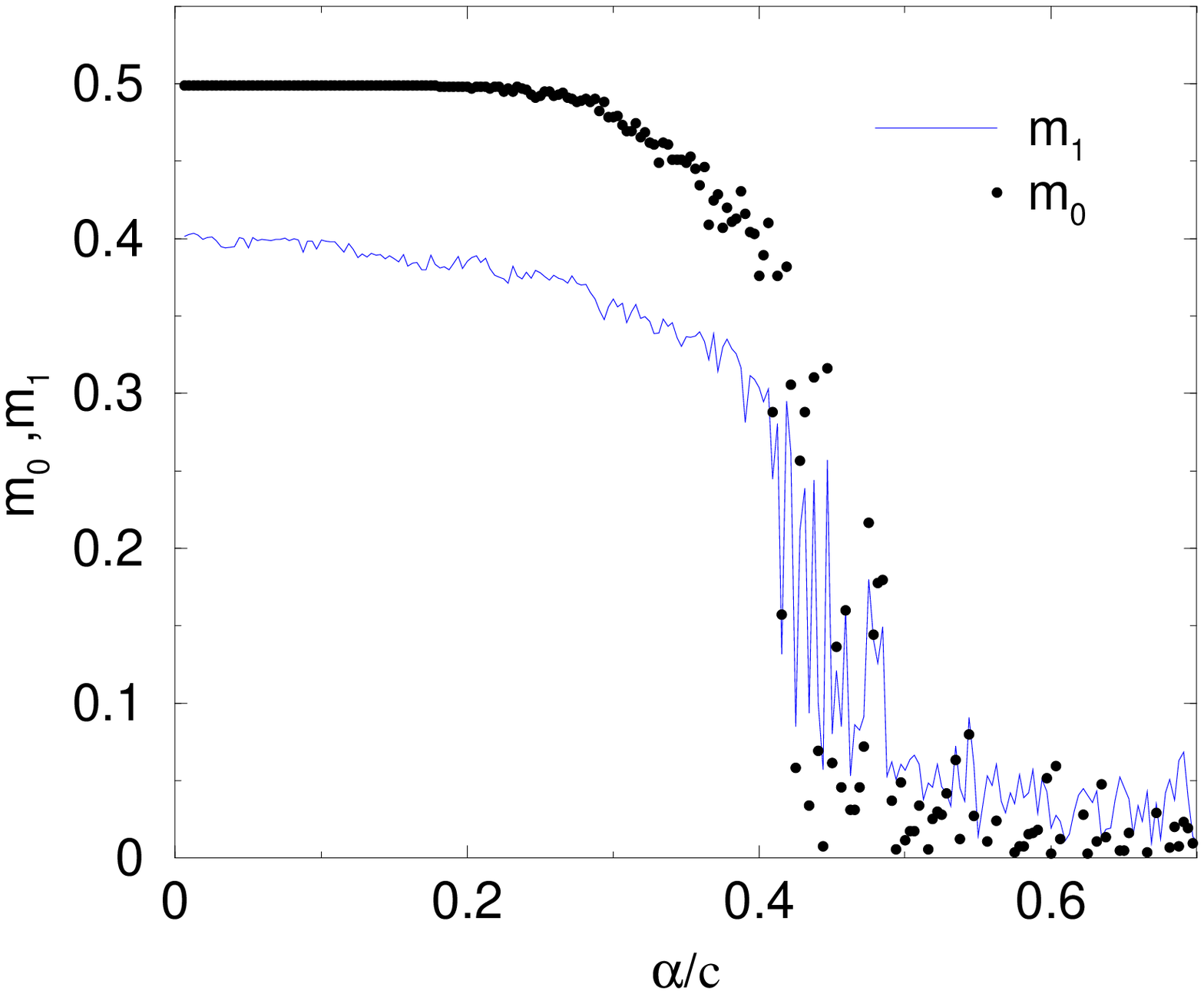}
\end{minipage}\ {\it(a)}\ \ \ \ \ \ \ \ \ 
&
\begin{minipage}{5cm}
\epsfxsize 5cm 
\epsfbox{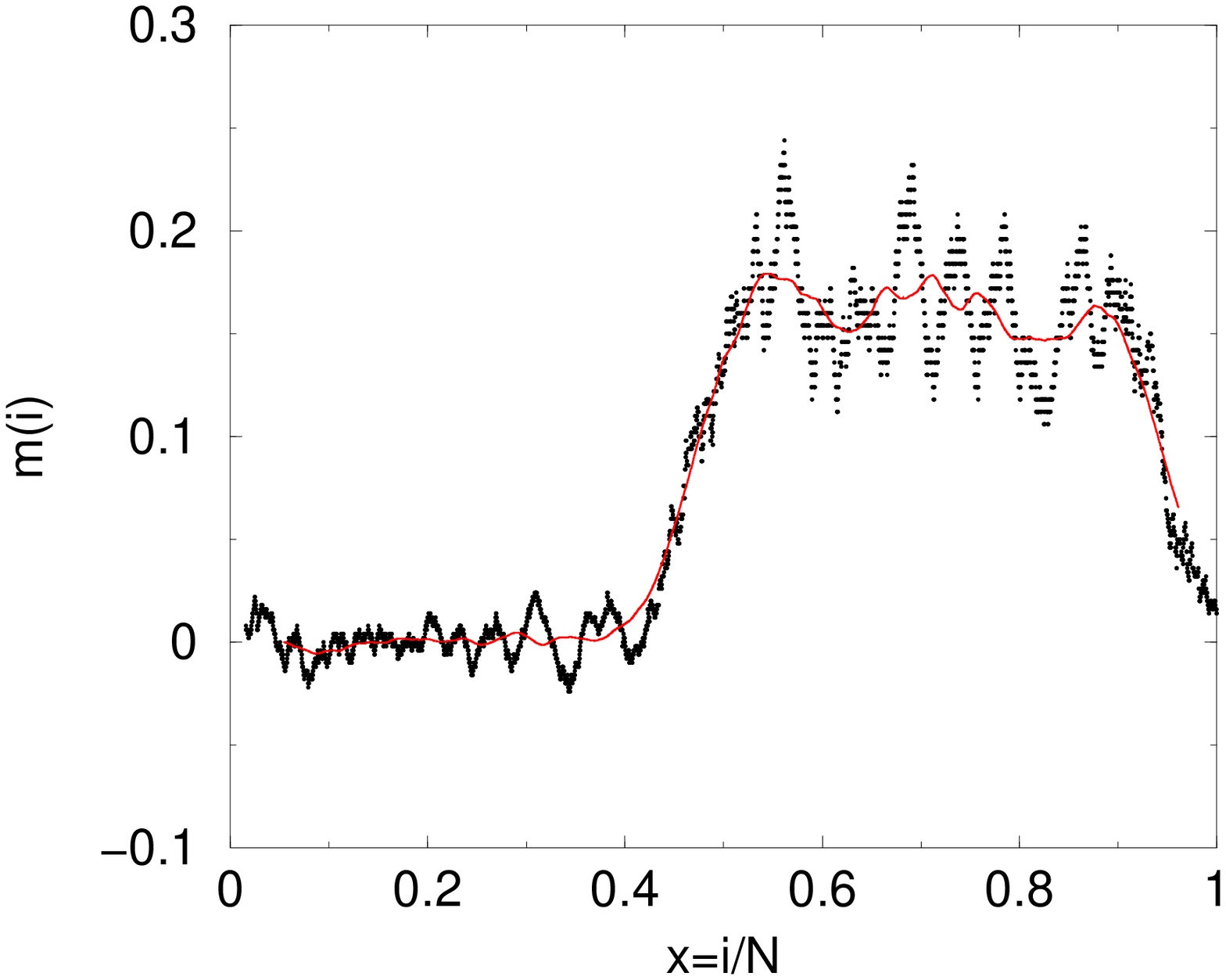}
\end{minipage}\ {\it(b)}
\\\\\\
\begin{minipage}{5cm}
\epsfxsize 5cm 
\epsfbox{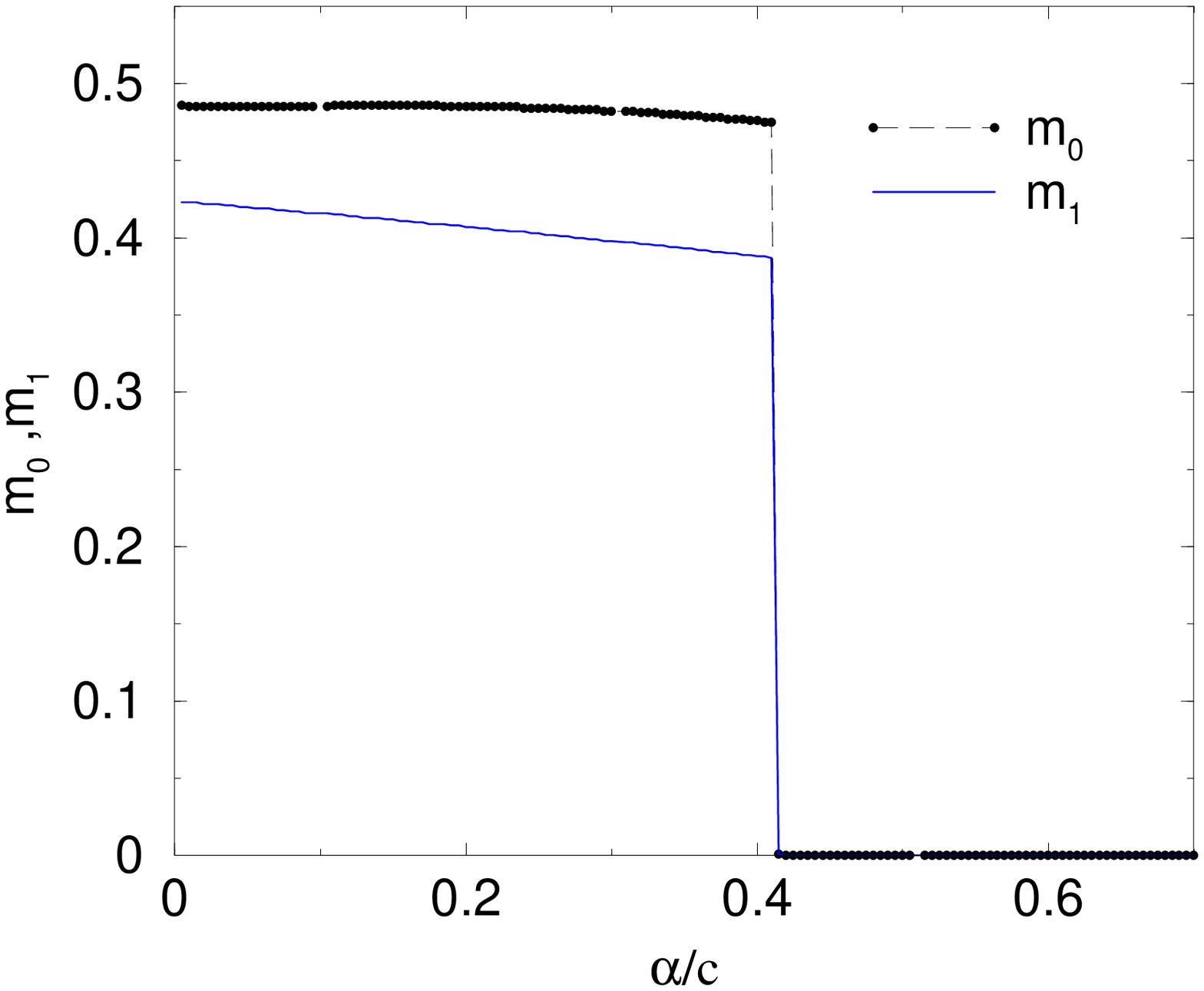}
\end{minipage}\ {\it(c)}\ \ \ \ \ \ \ \ \ 
&
\begin{minipage}{5cm}
\epsfxsize 5cm 
\epsfbox{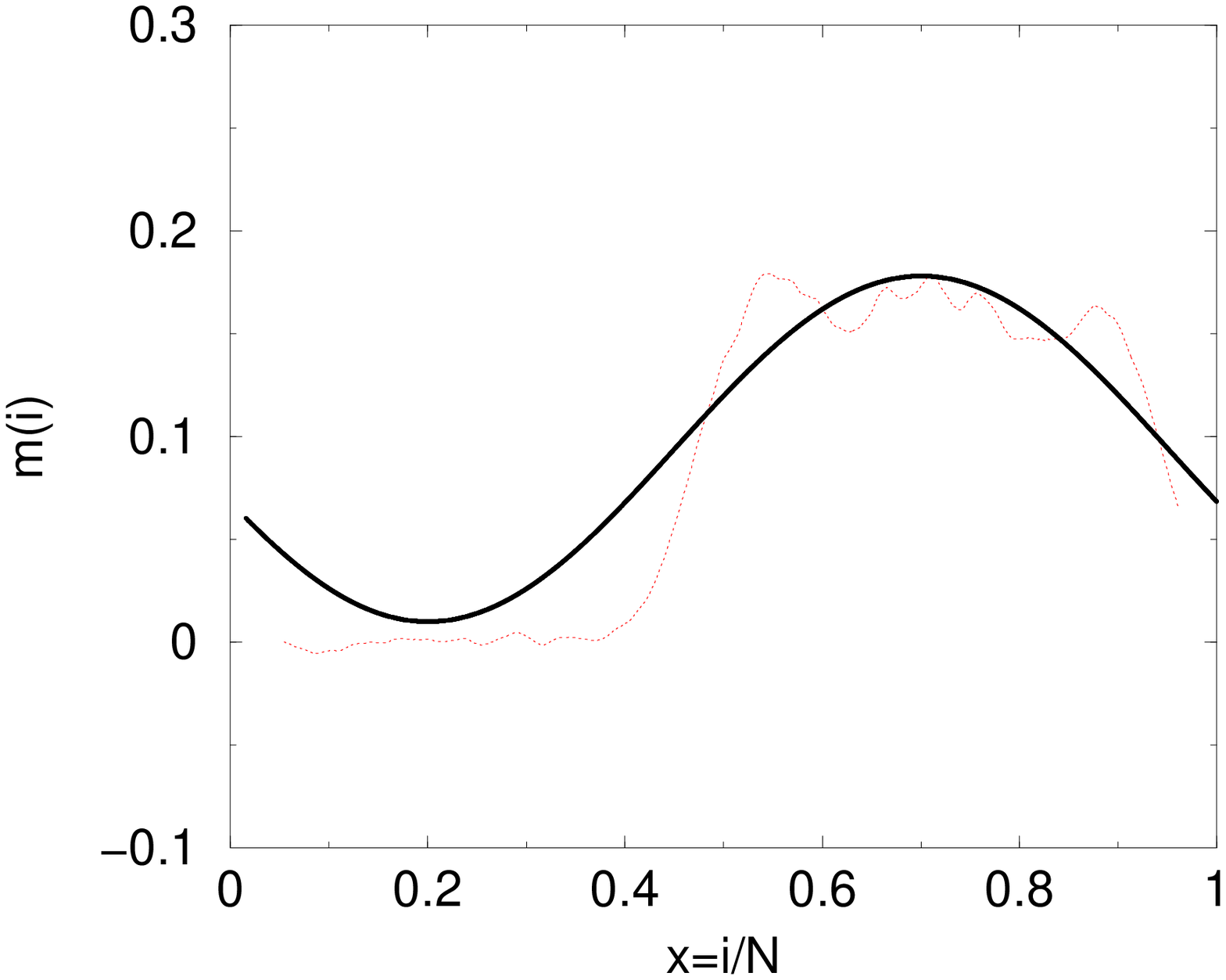}
\end{minipage}\ {\it(d)}
\end{array}$
\caption{\small
(Color online) 
Simulation (a)  and computation of $m_0,m_1$ according to
Eqs.(\ref{T=0}-\ref{tzeron}) (c).
The sparsity of the code is $a=0.8$ and $R=0.57$.
The form of the bump, represented by its local overlap, Eq.(\ref{loc_overlap}),
according to the simulation (b) and
the theory -- (d) bold line.
Parameters of the simulation: $p/(cN)=0.1$,
$N=6400$, $c=0.05$, $\sigma_x=500$, $m(i)$ smoothed with lengths 100 and 500; 
$\mu_1=0.95$.
}
\label{newnewfig}
\end{center}
\end{figure}
\end{widetext}
For values of the order parameters $m_1=m_2=...=0$ and $R=0$,
we obtain a result similar to that of Canning and Gardner \cite{Gardner}. 
The classical result of Amit et al \cite{Amit1} is obtained when $\mu_1=0$.

The result of the numerical solution of Eqs.(\ref{T=0}-\ref{tzeron})
is show in 
Fig.\ref{newnewfig}(c,d), 
where we have rescaled the
order parameters $m_0$ and $m_1$ to belong to the interval [0,1],
instead of the interval $[0,1-a]$, for any sparsity $a$.
The sharp bound of the phase transition is a result of taking into account
just two terms of $m_k$, $k=0,1$ and the lack of finite-size 
effects in the thermodynamic limit. 
The corresponding behavior of $m_0$ and $m_1$, 
obtained by simulation, is presented in the top panels of 
Fig. \ref{newnewfig}(a,b).
The good correspondence between the numerical result and the simulations
confirm our hypothesis of the finite number of macroscopic OP $m_l\neq 0$.
The physical
reason for this suggestion lies on the fact that the bump formations 
are compact structures with a large size and therefore the low-frequency 
components, e.g those corresponding to large eigenvalues of the matrix 
of connectivities, are responsible for the observed behavior.

\begin{figure}[t]
\begin{center}
\begin{minipage}{5cm}
\epsfxsize 5cm 
\epsfbox{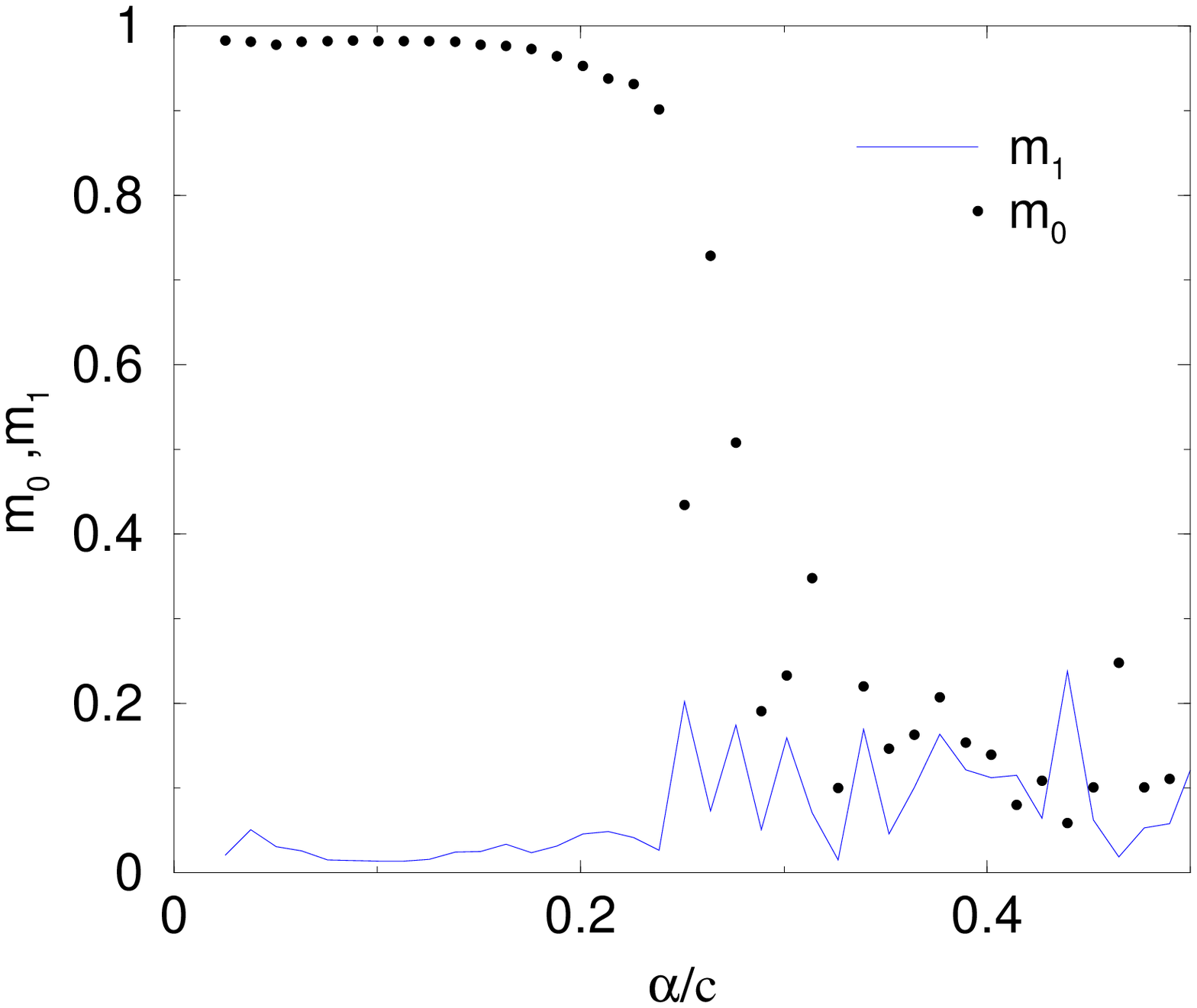}
\end{minipage}\ ($a$)
\vskip0.3cm
\begin{minipage}{5cm}
\epsfxsize 5cm 
\epsfbox{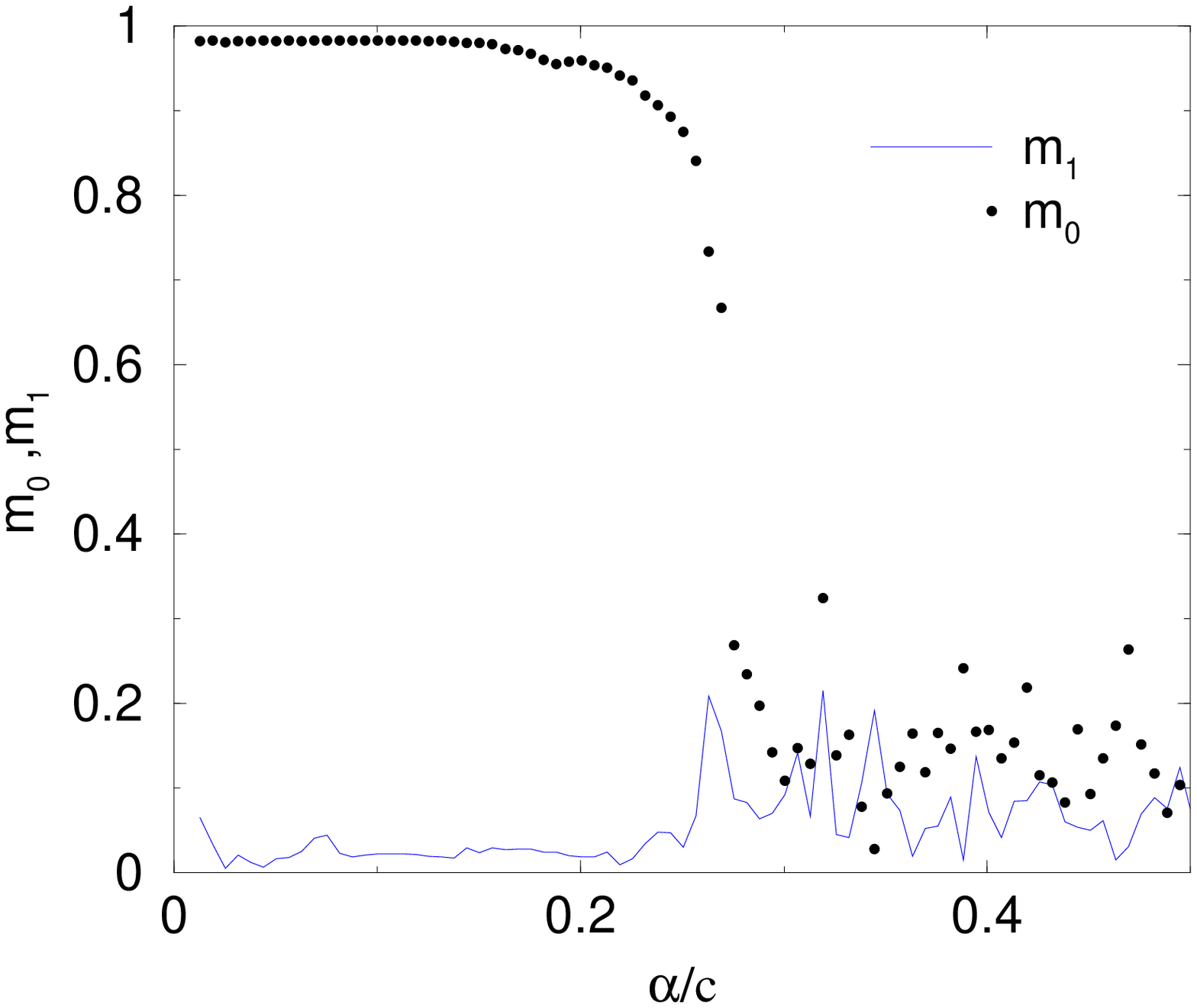}
\end{minipage}\ ($b$)
\caption{\small
(Color online) 
Binary attractor networks with the same distribution of the
pattern and the retrieval activity (R=0).
The overlap $m$ and the power of its first Fourier transform
are chosen as a measure of the existence of SAS.
The panel (a) is for
$N=6400$, $c=80/N$, $\sigma_x=100$, $a=0.8$, the panel (b) is for
$N=6400$, $c=160/N$, $\sigma_x=200$, $a=0.8$. None of the networks
present SAS.
The same is true for sparse code,
different dilution and other topologies as for example the defined by 
Eqs.(\ref{conxgauss},\ref{conxcos},\ref{conxsw}).
}
\label{fig1}
\end{center}

\end{figure}
\begin{figure}[t]
\begin{center}
\begin{minipage}{5cm}
\epsfxsize 5cm 
\epsfbox{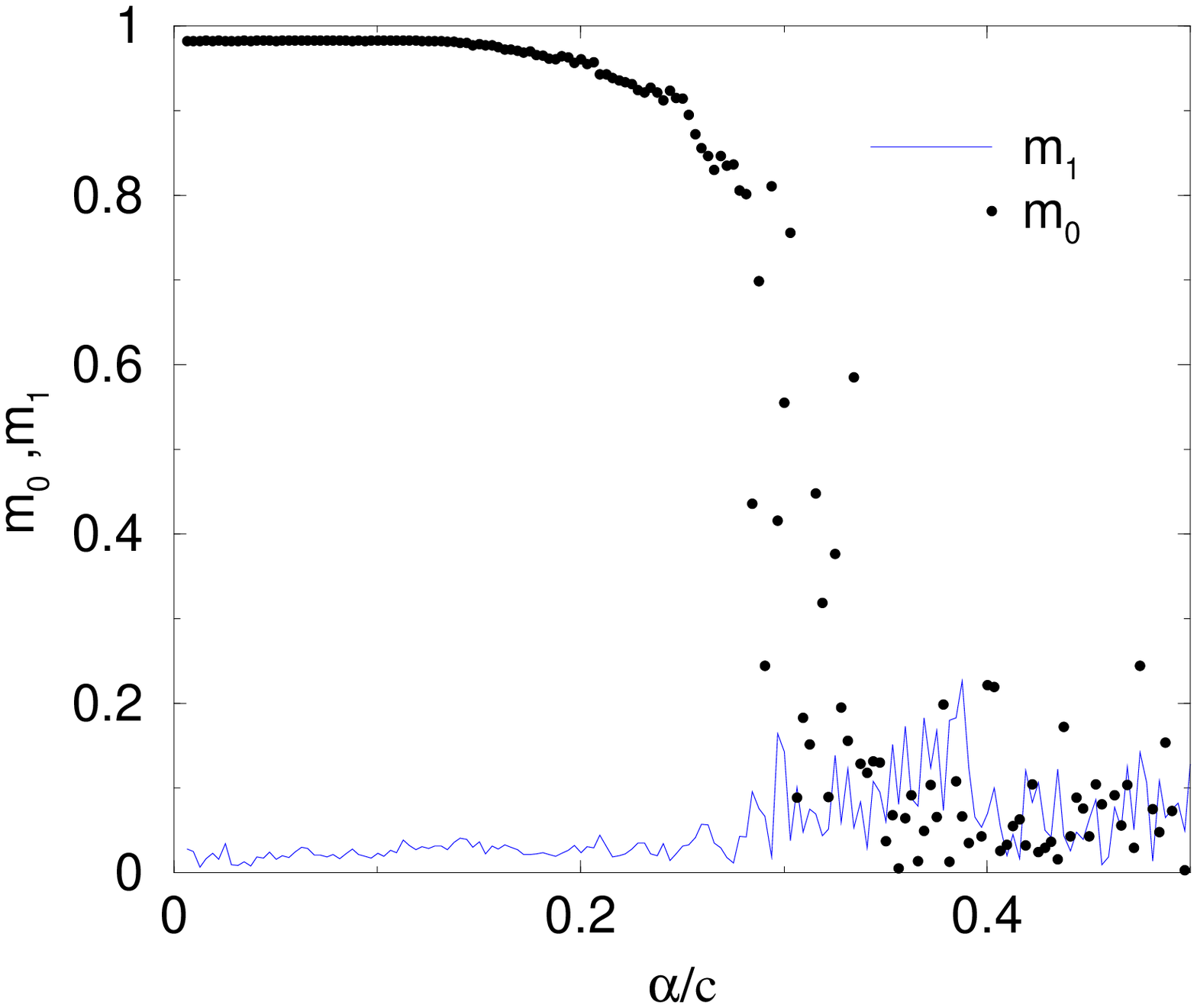}
\end{minipage}\ ($a$)
\vskip0.3cm
\begin{minipage}{5cm}
\epsfxsize 5cm 
\epsfbox{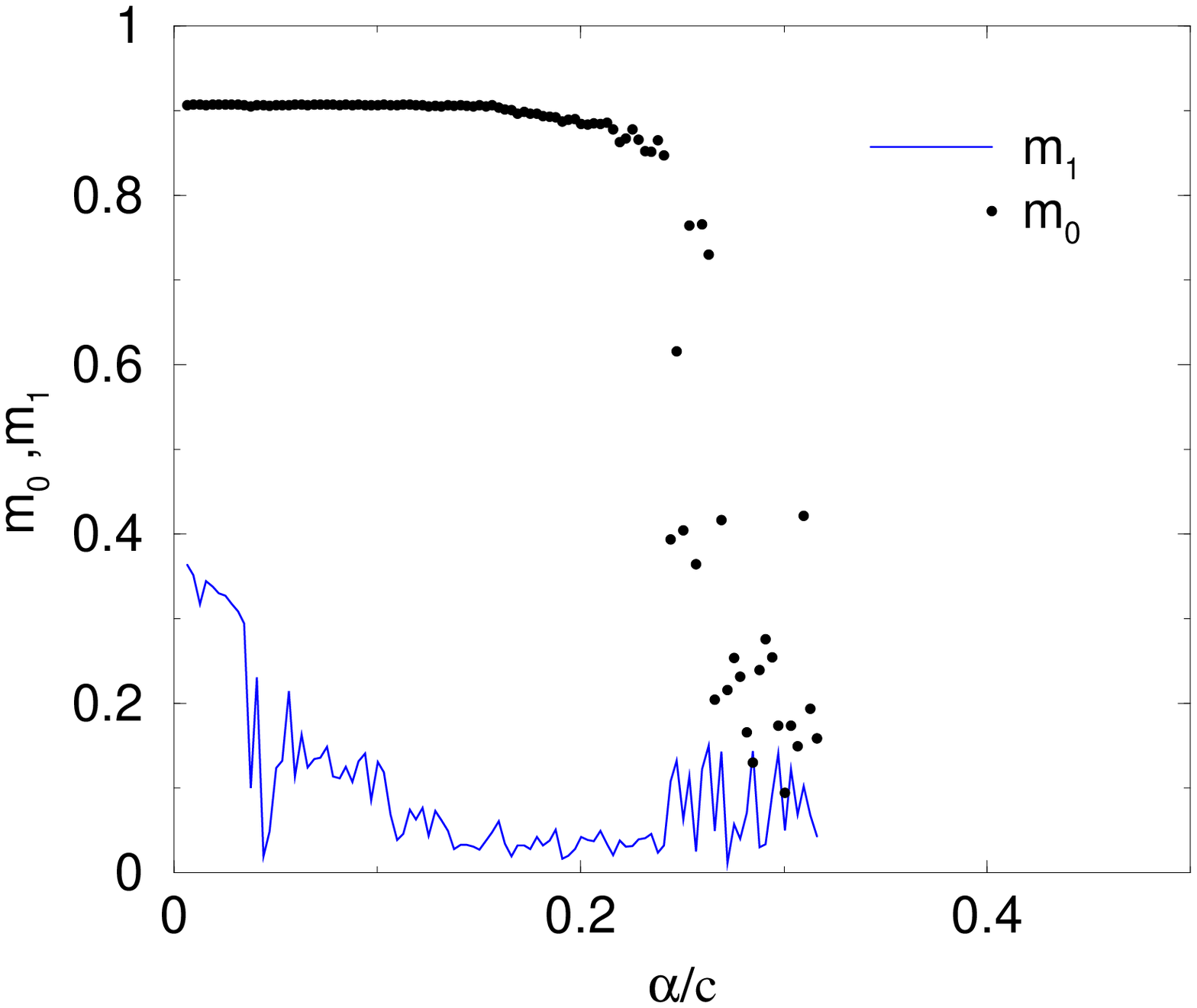}
\end{minipage}\ ($b$)
\caption{\small
(Color online) 
Spatial asymmetry observed by its first Fourier
component power $m_1$ with symmetric code ($a=0$).
The panel (a) shows the network with
$N=6400$, $c=0.05$, $\sigma_x=500$ and $R=0$.
The panel (b) shows the network with
$N=6400$, $c=0.05$, $\sigma_x=500$ and $R=0.5$.
SAS is observed only when $R\neq 0$.
}
\label{fig2}
\end{center}
\end{figure}

\section{Simulations}
We performed simulations for a network in form of 
a circular ring,
with a distance between the neurons $i$ and $j$:
\[
|i-j| \equiv \min(i-j+N\ {\rm mod}\ N,j-i+N\ {\rm mod}\ N),
\]
using three different topologies, explained in this section.

To compare with the results of Ref.\cite{AleYasser1}, we work with a typical
connectivity distance $\sigma_x N$ . This Gaussian-like distribution defines
the first topology we are studying.
\begin{equation}\label{conxgauss}
P(c_{ij}=1)= c \left[
\frac{1}{\sqrt{2\pi}\sigma_x N} e^{-(|i-j|/N)^2/2\sigma_x^2}+p_0\right],
\end{equation}
where $p_0$ is chosen to normalize the expression in the brackets.
When $\sigma_x$ is small enough, then spatial asymmetry is expected.

The dynamics of the network at time $t+1$ and temperature $T=0$ for our case
is
\[
S_i(t+1)={\rm sign}\left(\frac{1}{N} \sum_j \sum_{\mu} 
\xi_i^\mu \xi_j^\mu c_{ij} S_j(t) - T_h\right),
\]
where $T_h$ is the threshold of the system, which is in general nonzero,
due to the extra energy term $H_a$.

If the retrieval state, corresponding to the pattern $\xi_i^0$
is $S_i$, the mean overlap is $m_0=\sum_{i}\xi_i^0 S_i/N$ and
the local overlap at site $i$ is
\begin{equation}\label{loc_overlap}
m(i)\equiv\xi_i^0 S_i.
\end{equation}

In the case when $S_i$ follow a single sine wave,
the ideal measure of spatial asymmetry would be
\[
m_{1}=\frac{1}{N}\left|\sum_k \xi_k^0 S_k e^{2\pi i k/N}\right| ,
\]
which corresponds to the theoretically derived OP $m_1$ from 
Eqs.(\ref{m0},\ref{m}).

Simulations with more sharp localized
connectivity
\begin{equation}\label{conxcos}
P(c_{ij}=1) \propto \frac
{1-b \cos{\varphi} }
{1 - 2 b \cos{\varphi} + b^2 },
\end{equation}
with $\varphi\equiv 2\pi|i-j|/N$ and $b$ being some parameter, show
similar results. This connectivity defines the second type of topology
we are studying.
It has the advantage that the eigenvectors of the connectivity
matrix are cosine waves and the eigenvalues are known.
The results of the simulations
for different $\sigma$ and $a$ are shown in Figs. \ref{newnewfig}-\ref{fig2}.
These two topologies, Eqs.(\ref{conxgauss},\ref{conxcos}), 
give very similar results.

If $R=0$, no asymmetry can be observed for any $\sigma_x$,
up to the level of the network fragmentation (Fig. \ref{fig1}).
No difference between
asymmetric and symmetric connectivity
is observable for any connectivity
$c<0.05$ and any of the topologies tested.

The sparse code increases the SAS effects: Fig.\ref{fig2}(b),
vs. Fig.\ref{newnewfig}(a), 
but SAS can not be observed for any
sparsity $a$ if the proportion of the firing neurons is kept to be equal
to $a$ , i.e. when
the retrieval and the memorized states have the same level of activity.
However, even without sparsity ($a=0$) the bump states can be 
observed Fig.\ref{fig2}(b).

One could expect that the bump formations actually will deteriorate the performance of the neural networks, because the bump is localized in a limited part of the network and this effectively shrinks the active part of the network. 
However, it results not to be so. 
Namely, if we start the retrieval process from the bump $m_0 < 1$, fixing $R$ in such a way that a bump will be produced, this retrieval is better than the retrieval achieved, starting from totally overlapping pattern $m_0=1$. 

\begin{widetext}

\begin{figure}[t]
\begin{center}
$\begin{array}{ccc}
\begin{minipage}{4.70cm}
\epsfxsize 4.7cm 
\epsfbox{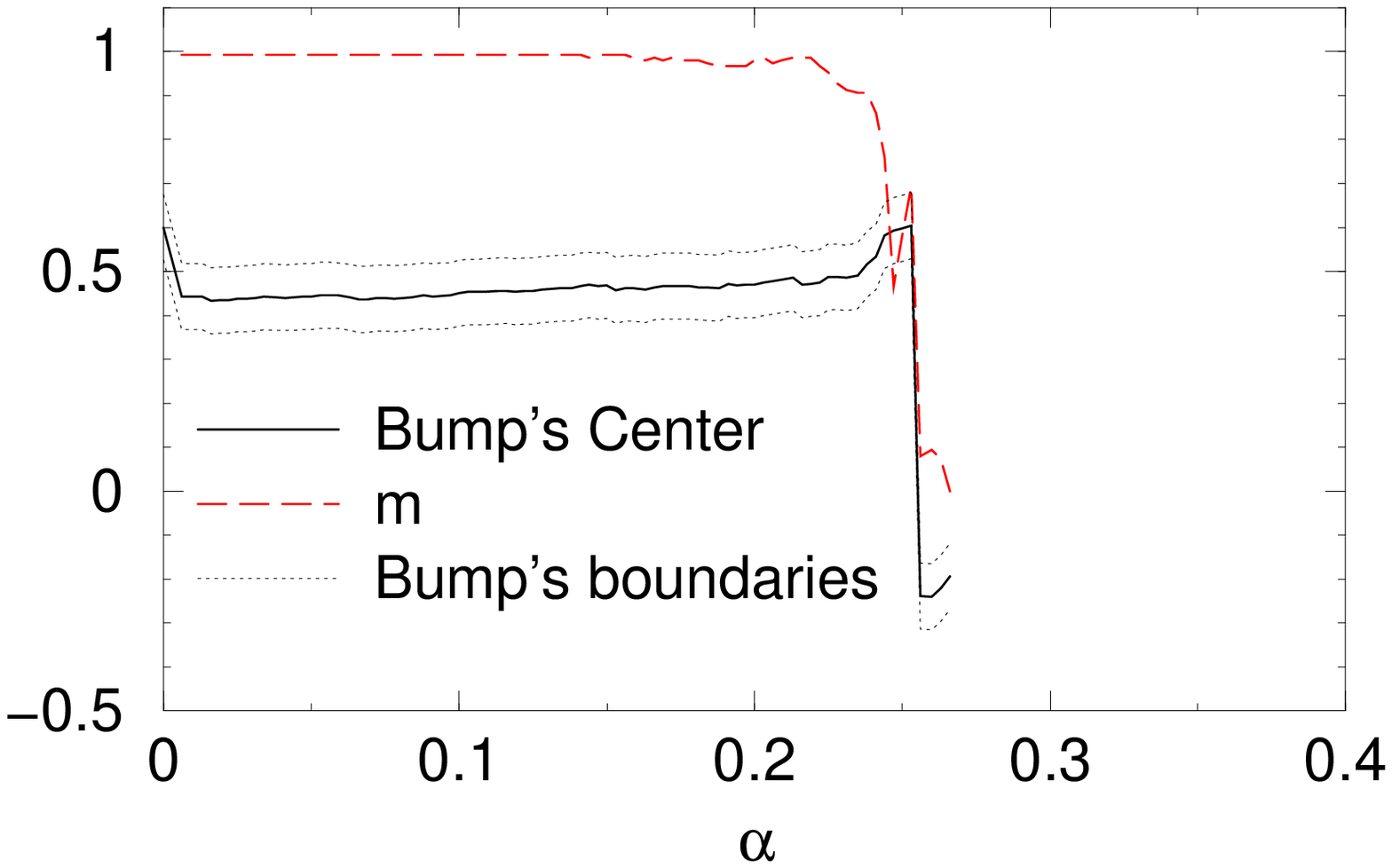}
\end{minipage}\ {\it(a)}
&
\begin{minipage}{4.70cm}
\epsfxsize 4.7cm 
\epsfbox{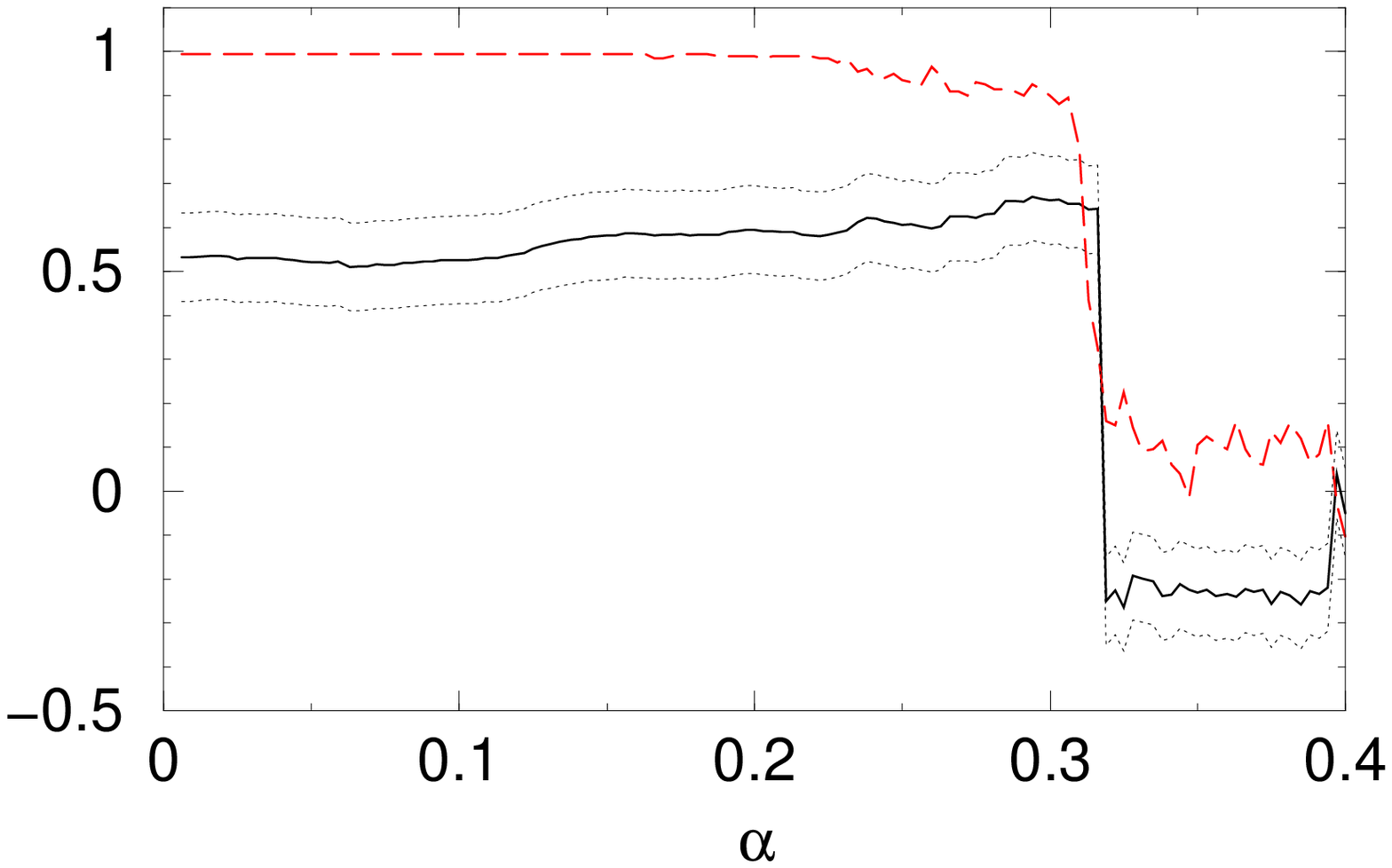}
\end{minipage}\ {\it(b)}
&
\begin{minipage}{4.70cm}
\epsfxsize 4.7cm 
\epsfbox{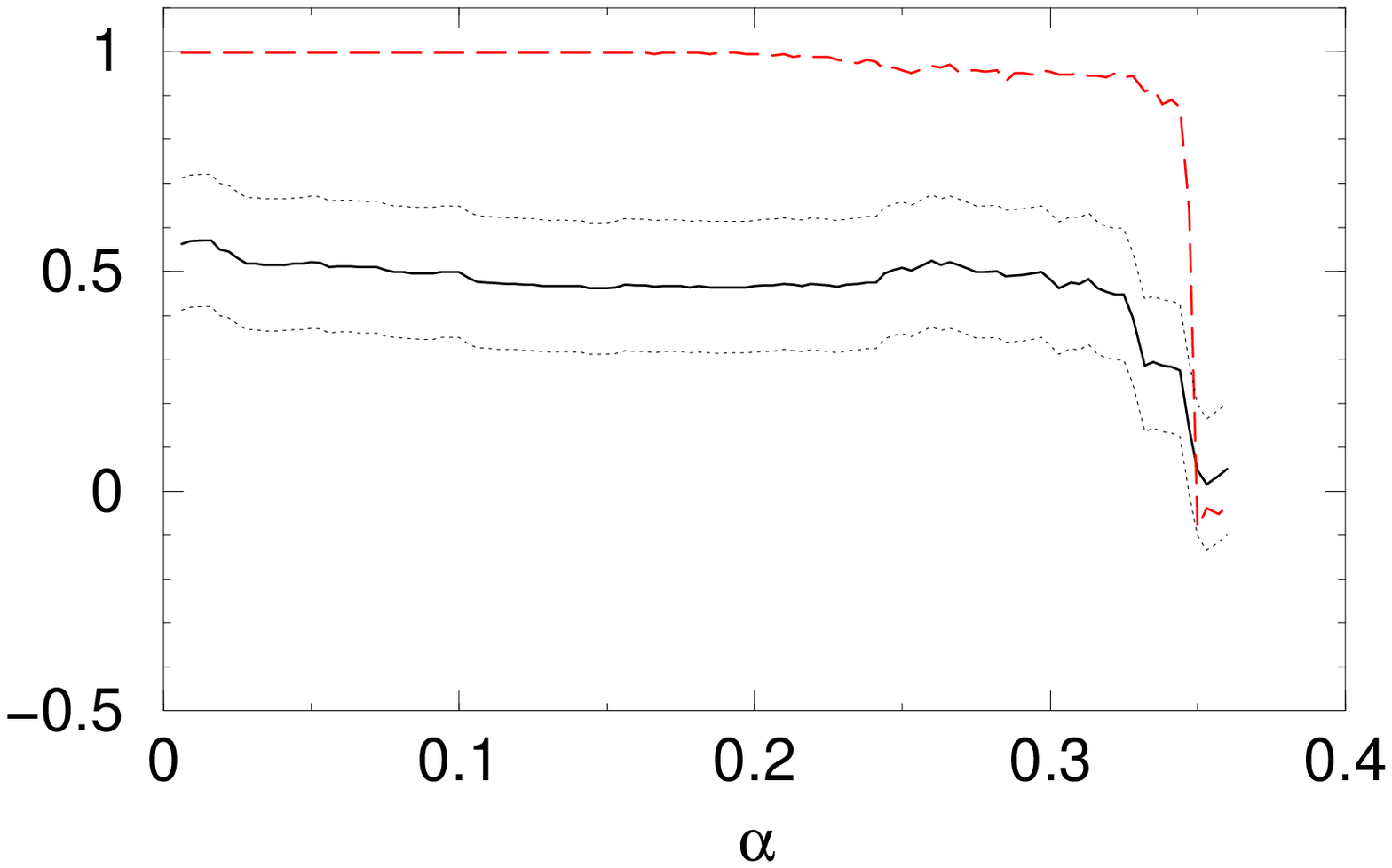}
\end{minipage}\ {\it(c)}\\
\begin{minipage}{4.70cm}
\epsfxsize 4.7cm 
\epsfbox{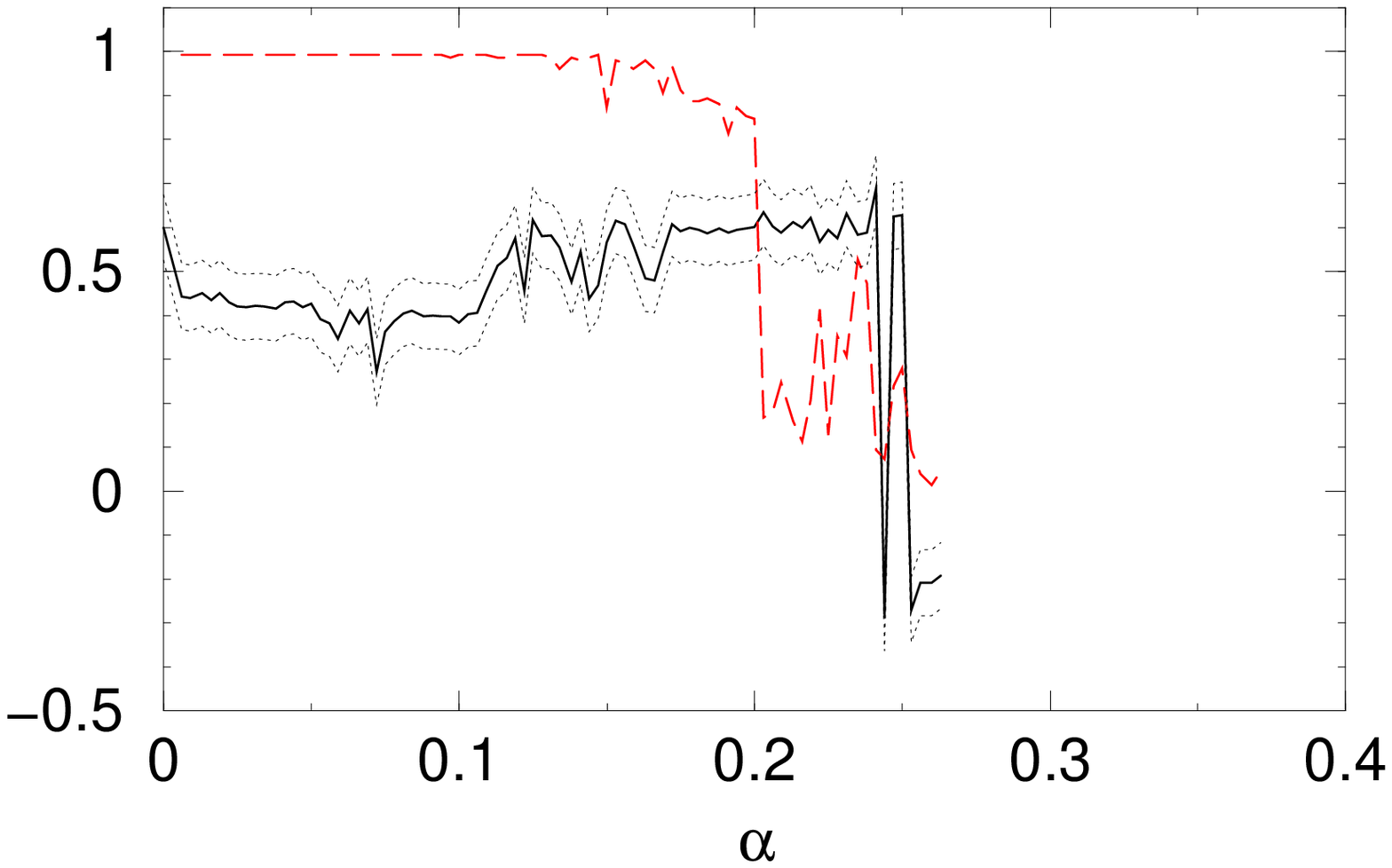}
\end{minipage}\ {\it(d)}
&
\begin{minipage}{4.70cm}
\epsfxsize 4.7cm 
\epsfbox{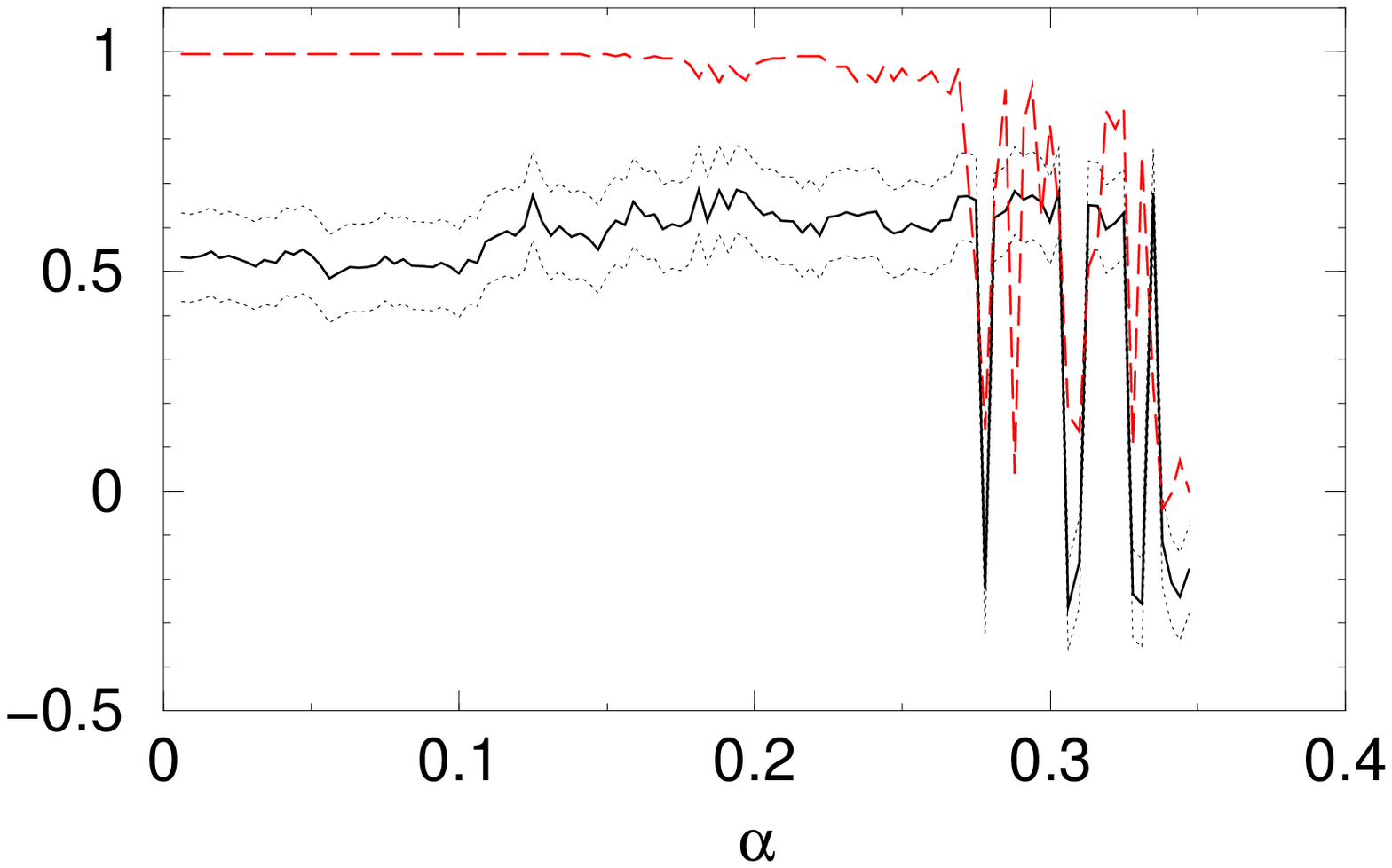}
\end{minipage}\ {\it(e)}
&
\begin{minipage}{4.70cm}
\epsfxsize 4.7cm 
\epsfbox{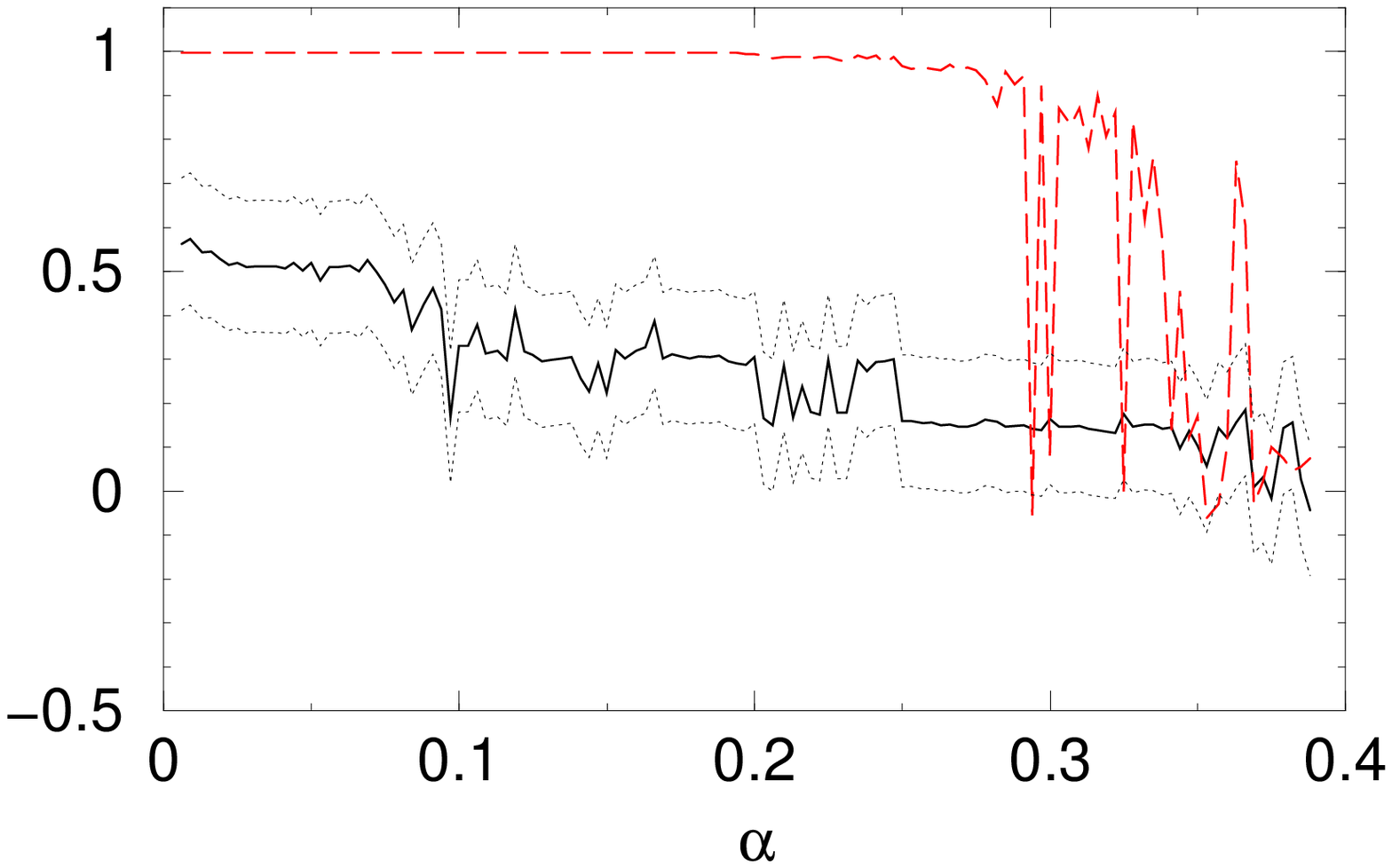}
\end{minipage}\ {\it(f)}
\end{array}$
\caption{\small
(Color online) 
  The dynamics of the bump formations. On the vertical axis the center 
and the boundaries of of the bump, as well as the asymptotic value of $m$ are given as a function of the 
load $\alpha$ for three different initial conditions $m_0=0.15, m_0=1$ 
(Fig.a and d), $m_0=0.2, m_0=1$ (Fig.b and e) and finally $m_0=0.3, m_0=1$ 
(Fig.c and f).
More stable behavior is observed when starting from the bump (Fig.a,b,c), 
than starting from uniform initial condition (the pattern itself) (Fig.d,e,f). 
The parameters of the simulations are $N=6400, c=0.05$ and $\sigma_x=500$.
}\vskip-0.5cm
\label{initial_conditions}
\end{center}
\end{figure}
\end{widetext}

 In the case of one-dimensional 
network given with Gaussian-like distribution of the connectivity on a 
circular ring, we observe the behavior, given in 
Fig.\ref{initial_conditions}.

In Fig.\ref{initial_conditions} we observe a
pronounced 
stability of the position of the center and the boundaries of the bump, 
when using the bump initial condition, 
while in the 
case of using the whole pattern as an initial condition, this is no longer 
true. In the last case, one observes a shrinking of the region where the 
SAS state is observed, given by the rapid decrease of the order parameter 
(red dashed line), which occurs for smaller values of the load compared 
to the case of initial bump attractor.

This behavior is repeated for three different values of the bump initial 
conditions $m_0=0.15, m_0=0.2$ and $m_0=0.3$, given in 
Figs.\ref{initial_conditions}(a,b,c). The corresponding behaviors when 
starting from the whole pattern are given in 
Figs.\ref{initial_conditions}(d,e,f).

Figs. \ref{initial_conditions} also show that the difference between 
the two behaviors is smaller  
when the size of the initial bump is smaller. For example, when an initial 
bump condition is $m_0=0.3$, then the region of stability of the solutions 
extends up to value of the load $\alpha=0.35$, while it decreases 
up to $\alpha=0.29$ for uniform initial condition. For $m_0=0.15$, 
these values are $\alpha=0.25$ and $\alpha=0.2$ respectively.

The last result clearly points out the role of the bump formations on the 
stability during the retrieval.

\begin{figure}[ht!]
\begin{center}
\begin{minipage}{7.70cm}
\epsfxsize 7.7cm 
\epsfbox{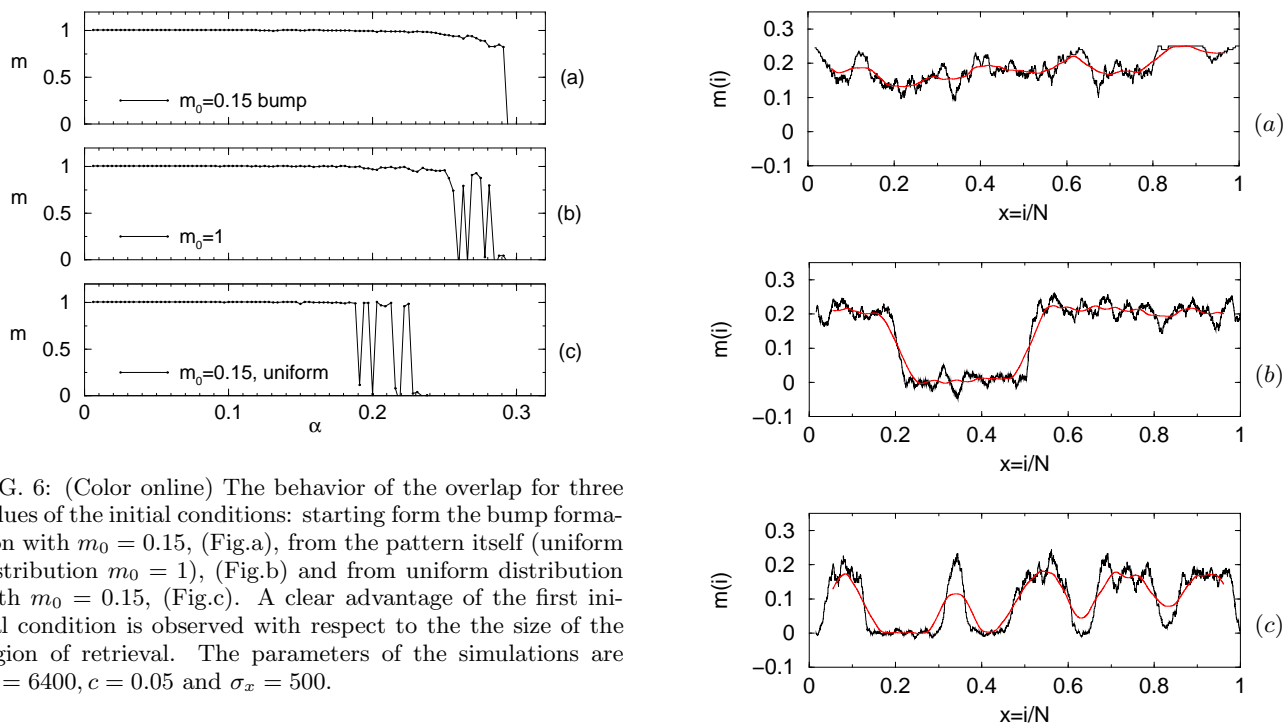}
\end{minipage}
\caption{\small
(Color online) 
The behavior of the overlap for three values of the initial conditions: 
starting form the bump formation with $m_0=0.15$, (Fig.a), from the pattern itself (uniform 
distribution $m_0=1$), (Fig.b) and from uniform distribution with $m_0=0.15$, 
(Fig.c). A clear advantage of the first initial condition is observed with 
respect to the the size of the region of retrieval. The parameters of the simulations are $N=6400, c=0.05$ and $\sigma_x=500$.}
\label{3cases}
\end{center}
\end{figure}

The comparison of the overlap in the case of initial conditions corresponding 
to
bump of size $m_0=0.15$, uniform distribution $m_0=1$ and uniform initial 
condition with $m_0=0.15$ is given in Fig.\ref{3cases}.

From Fig.\ref{3cases} it can be observed that the uniform initial 
condition with $m_0=0.15$ gives the worst result for the size of the 
retrieval solution. 
The overlap shrinks to zero for load less than $\alpha=0.2$, while this 
value is  $\alpha=0.29$ when starting form the bump. We face again with 
the conclusion that the bumps are stable formations and enhance the 
performance of the network.

This result could serve as a basis of the biological relevance of the 
bump formations for the whole stability of the network to learn and to 
retrieve information efficiently.

Finally, we also performed simulations in the case of the 
small-world (SW) 
topology \cite{SW1, SW2} 
with re-wiring rate $\omega\in [0,1]$:
\begin{equation}\label{conxsw}
P(c_{ij}=1)=(1-\omega)\theta(c-|i-j|/N)+\omega c,
\end{equation}
where $\theta$ is the theta function. For $R=0$, no SAS behavior is 
observed.

\begin{figure}[h!]
\begin{center}
\begin{minipage}{7cm}
\epsfysize3cm
\epsfbox{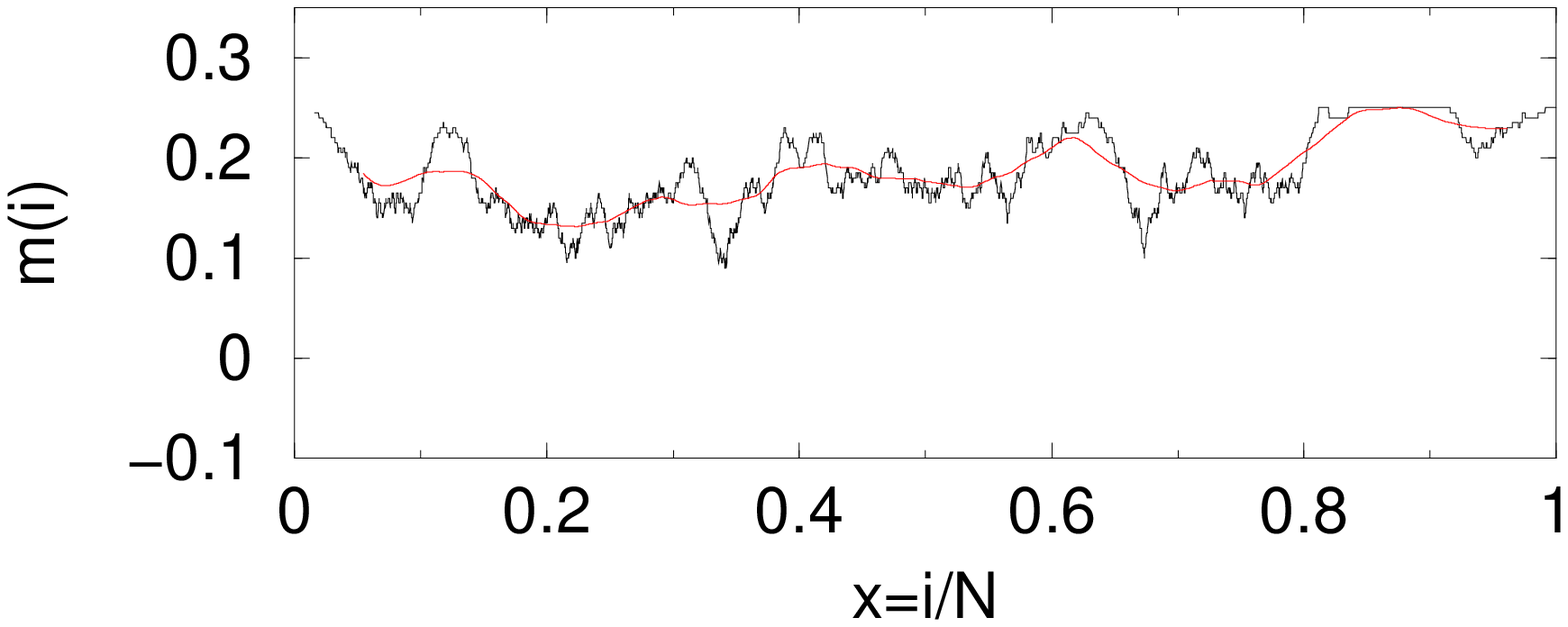}
\end{minipage}\ ($a$)
\vskip0.3cm
\begin{minipage}{7cm}
\epsfysize3cm
\epsfbox{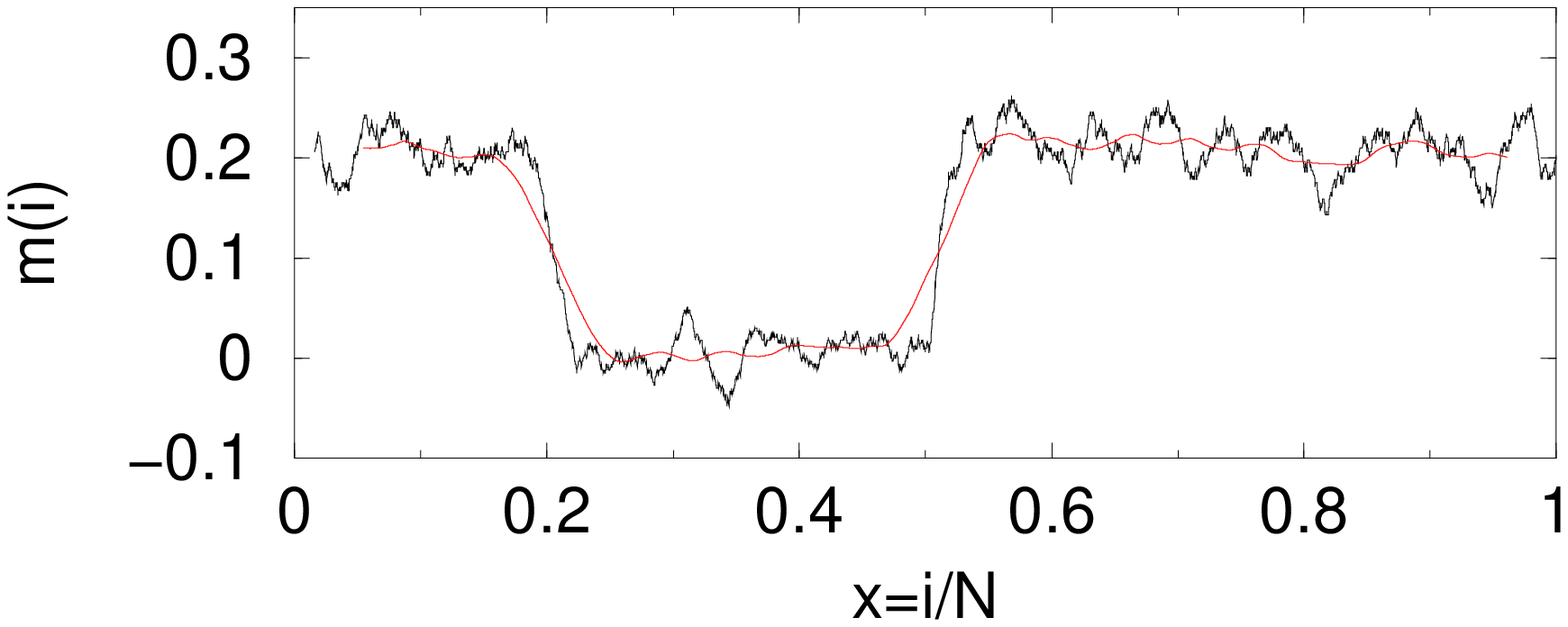}
\end{minipage}\ ($b$)
\vskip0.3cm
\begin{minipage}{7cm}
\epsfysize3cm
\epsfbox{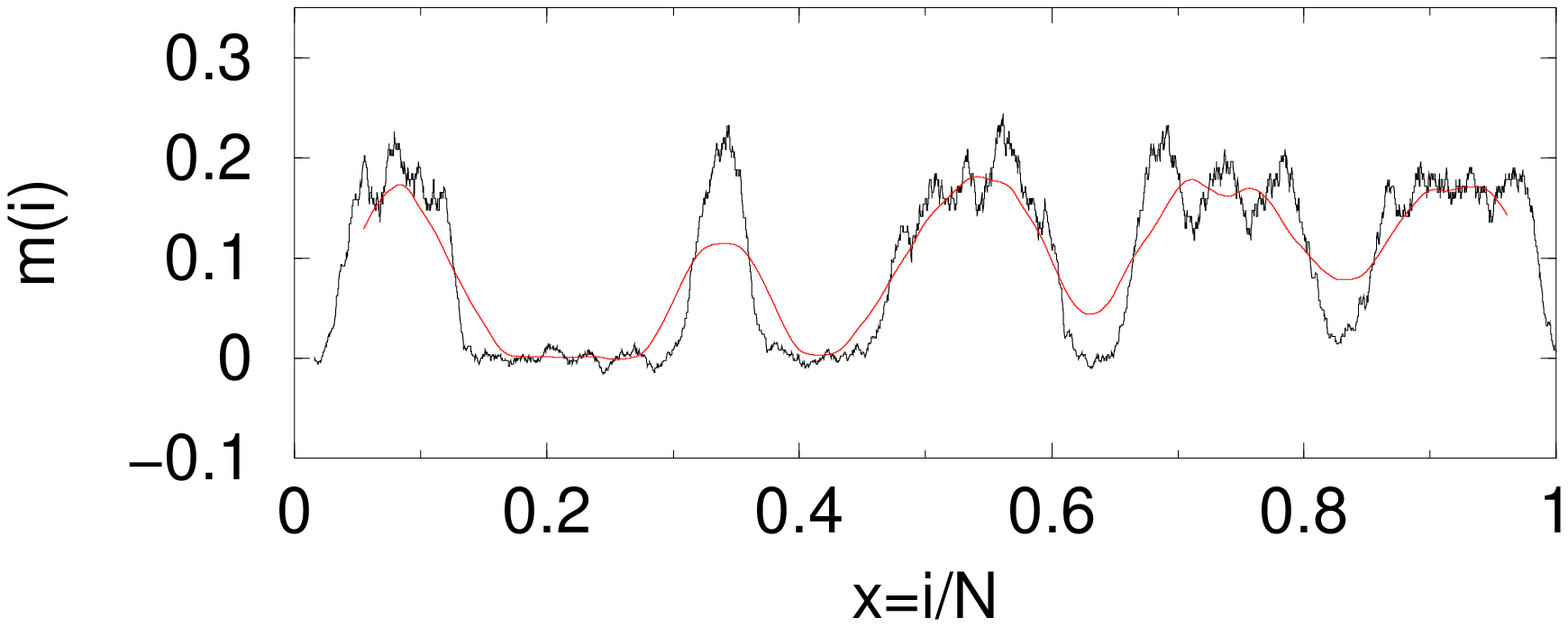}
\end{minipage}\ ($c$)
\caption{\small
(Color online) 
Smoothed local activity for small-world connectivity.
In the panel (a) no bumpiness, but significant fluctuations 
are present ($m_1=0.1 m_0$). In the panel (b) single bump is presented. 
And the last panel (c) shows multiple bump activity.
$N=6400$, $c=0.05$ and  $a=0.6$(a), $a=0.7$(b) and $a=0.8 $ (c), re-wire rate $\omega=0.06$, $R=0.5$. 
The same effect can be achieved by changing $(c,\omega)$.
}
\label{smallworld}
\end{center}
\end{figure}

However, when $R\ne 0$,
the simulations show more complex
behavior with the appearance of several bump formations,
as well as phases with no bumps but significant $m_1$, Fig.\ref{smallworld}.
It seems probable that
the multiple bumps are due to the roughness of the SW eigenvalues 
distribution \cite{kakvo}. This different behavior of the SW topology 
deserves future attention.

\section{Discussion}

\begin{figure}[t]
\begin{center}
\begin{minipage}{5.70cm}
\epsfxsize 5.7cm \epsfysize=5cm
\epsfbox{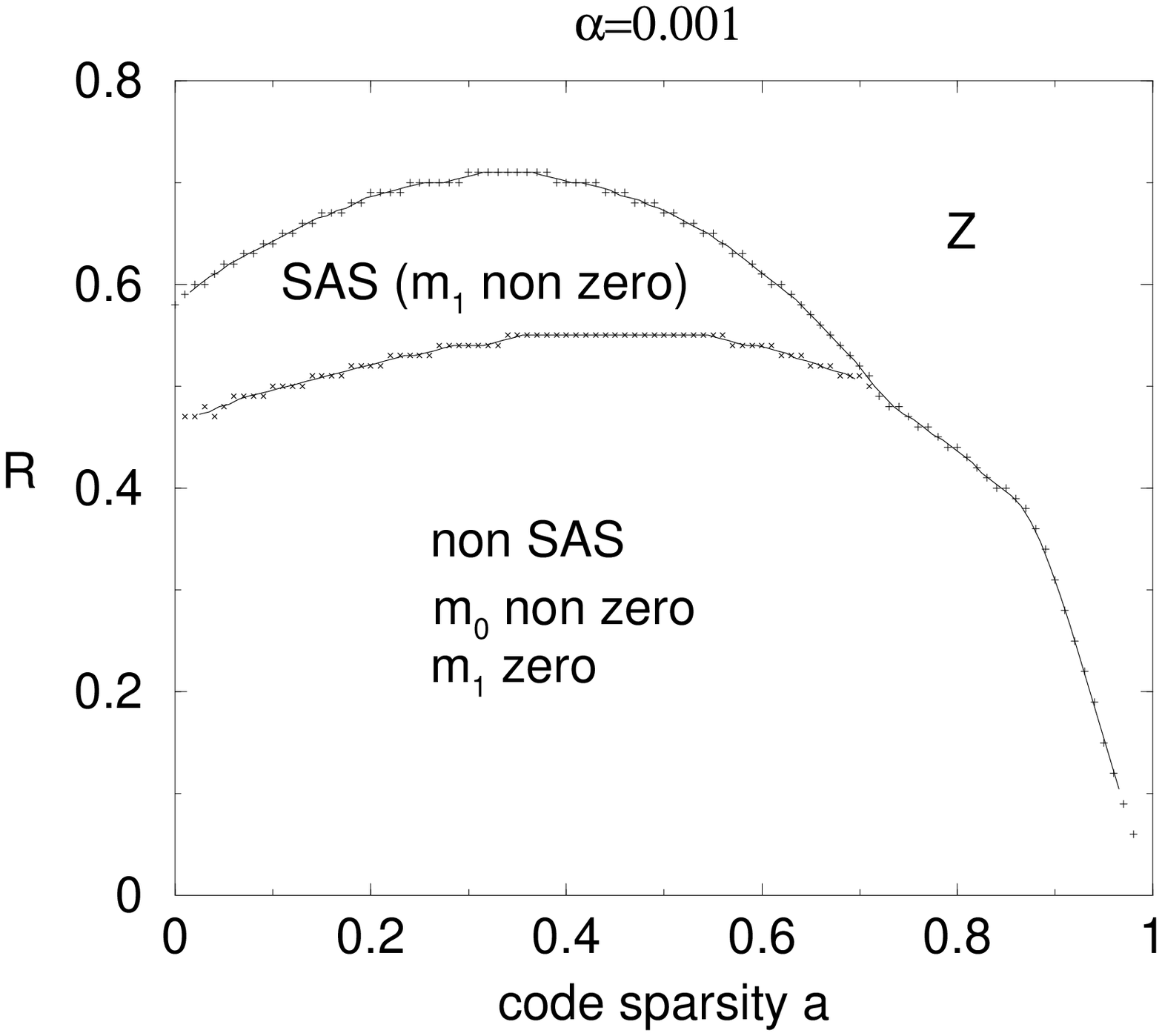}
\end{minipage}\ ($a$)
\vskip0.3cm
\begin{minipage}{5.70cm}
\epsfxsize 5.7cm 
\epsfbox{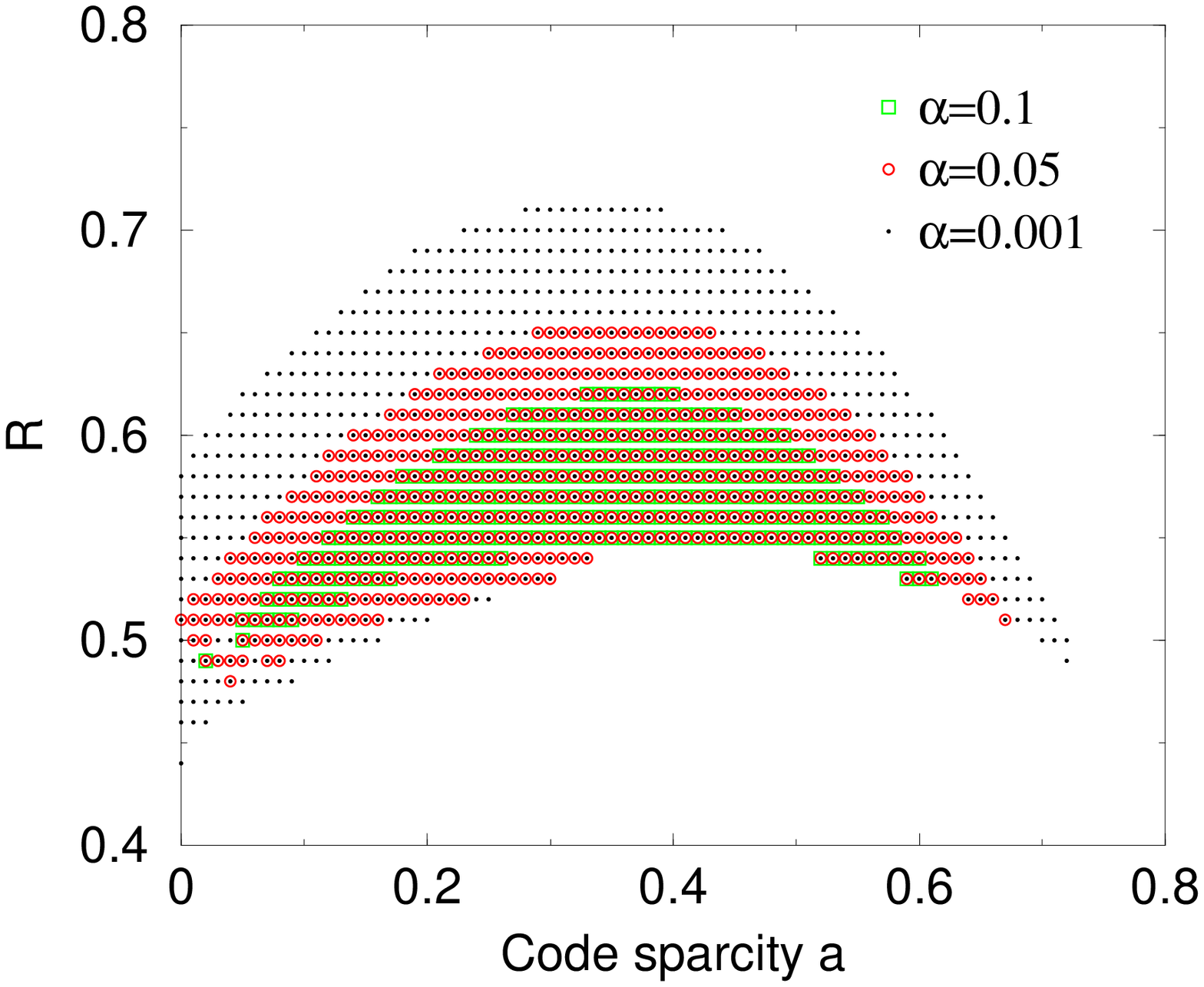}
\end{minipage}\ ($b$)
\caption{\small
(Color online) 
Phase diagram R versus $a$ for $\alpha=0.001$ ($\alpha/c=0.02$).
The SAS region, with $m_0 \neq 0, m_1 \neq 0$ is relatively large
and decreases by increasing the load (bottom).
Note that not all of the values of $\alpha$ are feasible with
arbitrary mean connectivity of the network $c$.
}
\label{phase}
\end{center}
\end{figure}

The numerical analysis of Eqs.(\ref{T=0}-\ref{tzeron}) gives 
a stable region for the
solutions corresponding to bump formations for different values of the load
$\alpha$, the sparsity $a$ and  the retrieval asymmetry parameter $R$, shown
in Fig.\ref{phase}. As can be seen, the sparsity of the code $a$ enhances the
SAS effect, although it is also observed for $a=0$. 

As we expected, the
asymmetry factor $R$  between the stored and retrieved patters is very
important in order to have spatial asymmetry.
The diagram in Fig.\ref{phase} shows a clear phase transition with $R$.
For small values of $R$, the effect (SAS) is not present.
If $R>1$, then the only stable state is the trivial $Z$ state,
as all the nonzero solutions are suppressed.
Only for intermediate values of $R$, the bump solution exists.

\begin{figure}[t]
\begin{center}
\begin{minipage}{5.70cm}
\epsfxsize 5.7cm\epsfysize4cm
\epsfbox{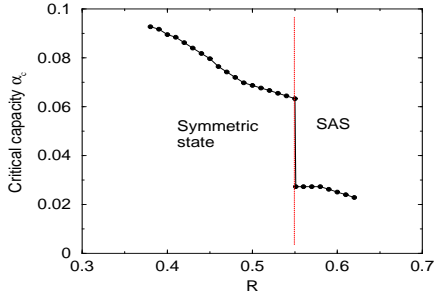}
\end{minipage}
\caption{\small
(Color online) 
The critical storage capacity $\alpha_c$ for $a=0.4$.
A drop of the capacity is observed on the transition between normal retrieval
state and state with localized activity.
}
\label{setcapacityR}
\end{center}
\end{figure}

When increasing the load, the area corresponding to SAS states shrinks
as is shown in 
Fig.\ref{phase}(b). 
This is due primary to the fact that 
the critical capacity $\alpha_c$ of the network drops when 
the system enters into the SAS state.
More precisely, the behavior of the critical storage capacity 
as a function of the asymmetry
parameter $R$, shown in Fig.\ref{setcapacityR}, presents 
a drastic drop of its value at the transition from 
homogeneous retrieval (symmetric) state to
spatially localized (asymmetric state).
Effectively, only the fraction of the network in the bump
can be excited and the storage
capacity drops proportionally to the size of the bump. For the values of the
sparsity of the code $a=0.4$, where the effect of the asymmetry parameter is
maximal, the decrease of the critical capacity is approximately twofold.


\begin{figure}[t]
\begin{center}
\begin{minipage}{8cm}
\epsfxsize 8cm 
\epsfbox{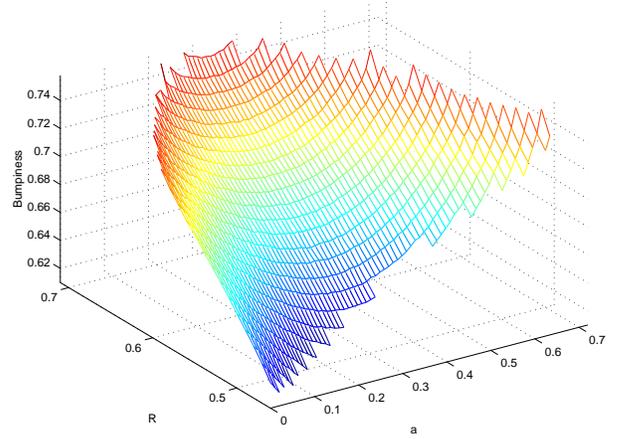}
\end{minipage}
\caption{\small
(Color online) 
The bumpiness $B$ of the network for different
values of $R$ and $a$.}
\label{bum}
\end{center}
\end{figure}

We have also investigated the effect of the parameters $R$ and $a$ 
on the bump formation. For this, we introduce a parameter that can 
give a criterion for
the bumpiness as the relative ratio between the two order 
parameters $m_0$ and $m_1$, given by Eqs.(\ref{T=0}-\ref{tzeron}):
\[
B=\sqrt{\frac{{m_{1}}^{2}}{{m_{0}}^{2}+{m_{1}}^{2}}}.
\]
The behavior of $B$ is presented in Fig.\ref{bum}.
It is seen that the simultaneous increase of both $R$ and $a$ inside the
area where the SAS behavior is observed, favors the bumpiness in the
binary network.
This behavior is due to the stronger localization of the activity 
as the extra term $H_a$ favors states with lower total activity. 
From the other side, increasing $a$ makes the code sparser, which favors the
metrical organization of both patterns and neurons.

Finally we have investigated the information capacity of the network.
For this aim we calculated the corresponding expression for the mutual
information \cite{Blahut} in the case of order parameters $m_0$ and $m_1$:

\begin{eqnarray}
I_M(m_0,m_1,\alpha) =
\frac{\alpha(m_0+1)}{2}\ln (m_0+1)\nonumber\\
+\frac{\alpha|m_1|}{\pi}F\left[\left\{\frac{1}{2},1,1\right\},
\left\{\frac{3}{2},\frac{3}{2}\right\},\frac{{m_1}^2}
{(1+m_0)^2}\right]\nonumber\\
+\frac{\alpha (m_0+1)}{2} \ln \left[\frac{1}{2}+\frac{1}{2}
\sqrt{1-\frac{{m_1}^2}{(1+m_0)^2}}\right].
\end{eqnarray}
Here $F[.]$ is the hyper geometric function \cite{abramz}.
The first term is the normal expression for the mutual information 
(the sparsity of the code eliminates the symmetry between $m$ and $-m$).
The next two lines of the equation can be
regarded as a correction to the usual expression due to OP $m_1\ne 0$.

\begin{figure}[t]
\begin{center}
\begin{minipage}{6.5cm}
\epsfxsize 6.5cm \epsfysize=5cm
\epsfbox{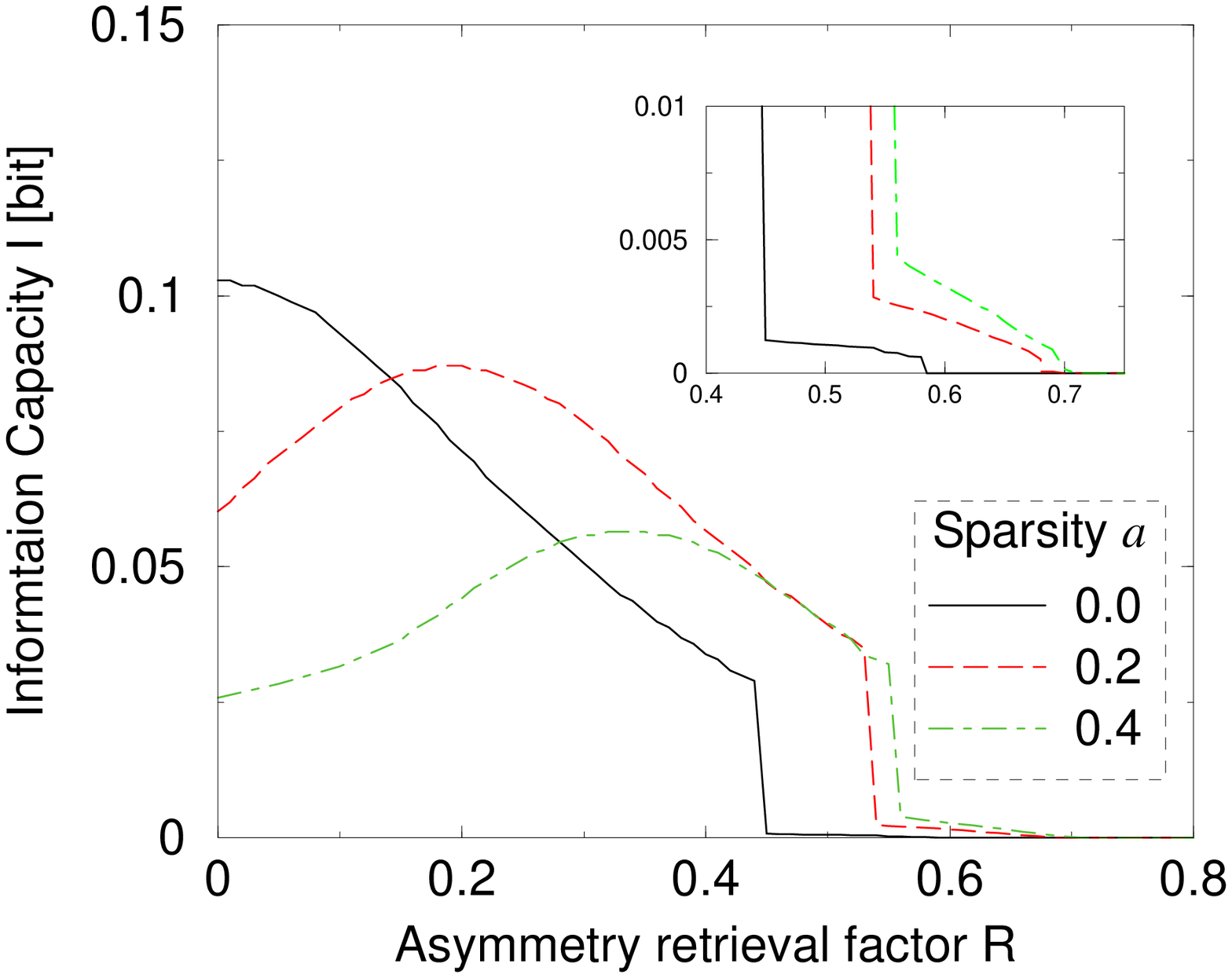}
\end{minipage}\ ($a$)
\vskip0.3cm
\begin{minipage}{6.5cm}
\epsfxsize 6.5cm \epsfysize=5cm
\epsfbox{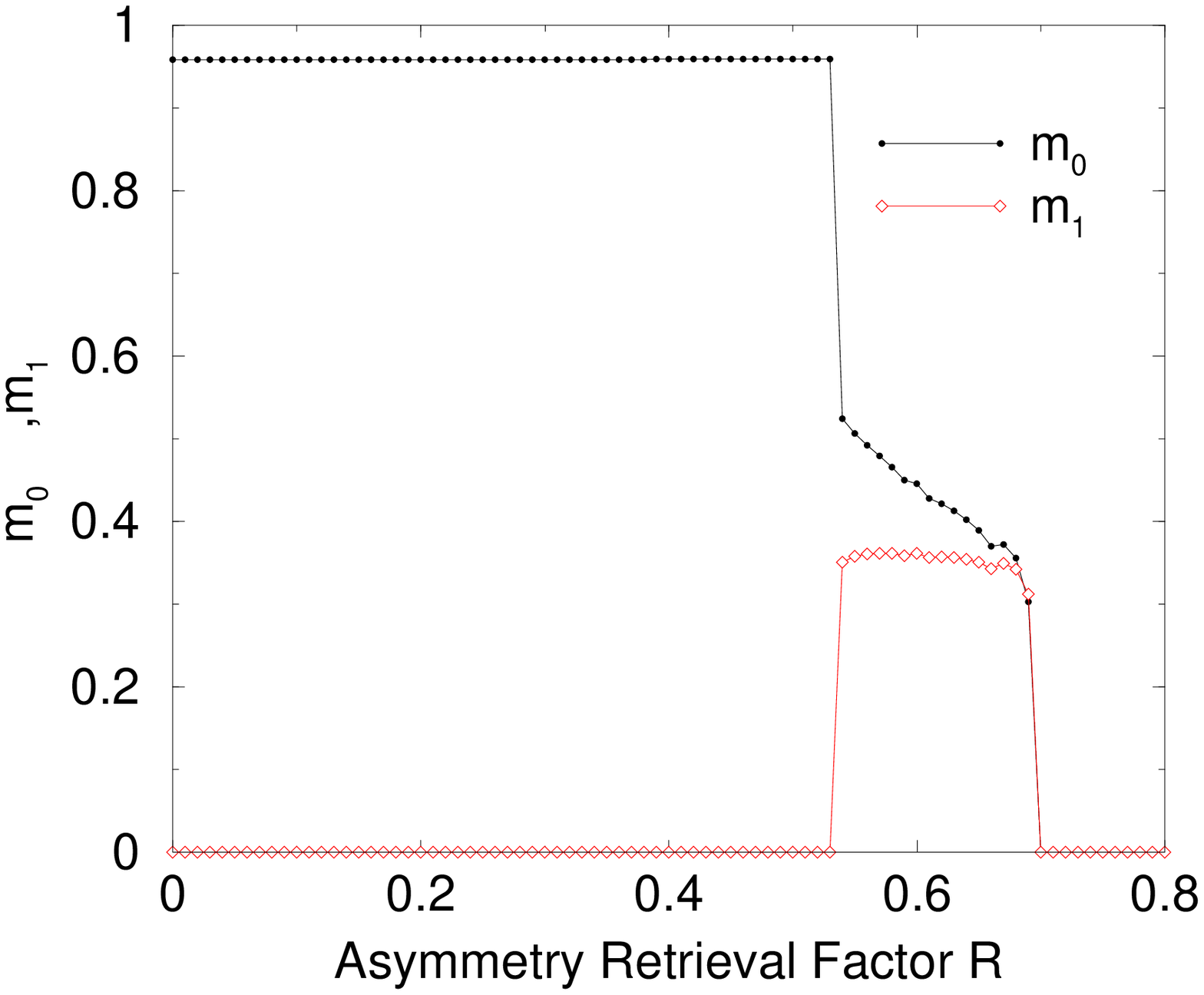}
\end{minipage}\ ($b$)
\caption{\small
(Color online) 
(a) The information capacity versus the asymmetry retrieval factor
$R$ for different values of the sparsity $a$. (b) $m_0, m_1$ for $a=0.2$. 
The inside figure in (a) shows the behavior of the information capacity near the point where it sharply decreases to zero.
}
\label{last}
\end{center}
\end{figure}

With the aim of Eqs.(\ref{T=0}-\ref{tzeron}) and the above expression, 
for the information
capacity $I_M/cN$ we obtained the behavior, presented in
Fig.\ref{last}, which has been also reproduced by simulations.
This behavior is non trivial and
shows a well pronounced maximum for intermediate values of the
sparsity of the code $a$. 
The corresponding behavior of the order parameters $m_0$ and $m_1$,
presented in Fig.\ref{last}(b), shows that inside the SAS
region, their values are significantly different form zero, but the
information capacity is very small although non zero.
If from one side, the information capacity $I_M/cN$, 
measures the number of bits,
needed to represent the state and if from the other side, 
the overlap parameters $(m_0, m_1)$ measure the degree to 
which one learned state can be recovered, then having 
small information capacity and large OP $m_0, m_1$ means that 
the original state can be recovered well with very small amount of
information.
This result could be useful for effective retrieval of patterns by
  using very small amount of information, especially in combination with the 
observed improvement of the stability of the network, 
provided that the retrieval starts from the bump.

 Similar results have been reported in
  \cite{AleYasser2} with respect to the increase of the computational power of
  the structured network. The last is expressed by the ability of the network
  to retrieve several patterns, each in a different location, whose
  combination could form the global combinatorially large pattern.

Apart of the advantage of the localized states in terms of the minimal
cost in transferring information, such structures could have important
biological relevance for the cortical modules of the mammalian brain, which
are characterized by geometrical distribution of the synaptic
connections. Recent experiments of the activity of a macaque temporal cortex
\cite{Rolls} have demonstrated that the receptive fields of the visual evoked
activity patterns are restricted up to three fold, when several patterns are
present together. The last result seems to be coherent with the picture of
localized activity in a neural network with metric organization of the 
synaptic activities.

\section{Conclusion}

In this paper we have studied the spatially dependent activity in
a binary neural network model when retrieval and learning states have
certain degree of asymmetry. We have shown that nor asymmetry of the
connections, neither sparsity of the code are
necessary to achieve bump solutions for the activity of the network, 
but rather
different symmetry between
the retrieval and the learning states.

The extension of the classical method of Amit et al. \cite{Amit1} in the case
of spatially dependent connectivity has permitted, within some
approximations, the derivation of the
above results analytically.
The good correspondence
with simulations, tested for different topologies of the network, is also
discussed through the paper. 



We have shown that the region of the spatially asymmetric states is
relatively large in the parameter space $(R, a)$ and we have measured 
quantitatively how the bumpiness of the network
depends on these two parameters.

A drop of the critical storage capacity is observed in the transition
form symmetric to SAS states, which is due to the fact that  effectively
only the fraction of the network in the bump
can be excited and the storage
capacity drops proportionally to the size of the bump.

When the recovering of the pattern starts from a state 
corresponding 
to the bump formation, the network behaves in a better way in terms of 
stability and capacity, compared to the case of uniform initial conditions. 
This happens even if the overlap with the uniform initial conditions is 
significantly larger.
The latter result could 
argue the biological relevance of the bump formations for effective 
retrieval of information.

Finally we have discussed the behavior of the information capacity
versus the asymmetry retrieval factor $R$ and we have discussed its sharp
decrease in the parametric region, where bump formations are present.

\acknowledgments 
The authors thank A.Treves and Y.Roudi for stimulating discussions at the
early stage of work as well as the Abdus Salam Center for 
Theoretical Physics for the financial support at the beginning of the  
present investigation.
E.K. and 
K.K. also acknowledge the
financial support from the Spanish Grants
DGI.M.CyT.FIS2005-1729
and 
TIN 2004--07676-G01-01 respectively.

\end{document}